\documentclass{article}
\usepackage[utf8]{inputenc}
\usepackage[margin=1in]{geometry}

\usepackage{xcolor}
\usepackage{graphicx}
\usepackage{amsmath}
\usepackage{amssymb}
\usepackage{cleveref}
\usepackage{csquotes}
\usepackage{siunitx}
\usepackage{bm}
\usepackage{authblk}
\usepackage{appendix}
\usepackage{booktabs}

\renewcommand{\vec}[1]{\bm{#1}}

\title{Numerical and  diagnostic extensions to the gyrokinetic \(\delta \! f\) code \texttt{stella}}
\author[1]{Sven Stroteich}
\author[2]{Georgia Acton}
\author[3]{Michael Barnes}
\author[2]{Pavlos Xanthopoulos}
\author[1]{Ralf Schneider}

\affil[1]{Institute of Physics, University of Greifswald, 17489 Greifswald, Germany}
\affil[2]{Max Planck Institute for Plasma Physics, Wendelsteinstr. 1, 17491 Greifswald, Germany}
\affil[3]{Rudolf Peierls Centre, University of Oxford, Oxford, United Kingdom of Great Britain and Northern
Ireland}

\date{\today}

\begin{document}

\maketitle

\begin{abstract}
Understanding turbulence in magnetised plasmas requires robust numerical tools capable of capturing complex free-energy transfer processes across scales. We demonstrate the utility of free energy as a diagnostic for both physical turbulence characteristics and numerical stability in gyrokinetic simulations with the \texttt{stella} code. Applied to the W7-X high-mirror configuration, free-energy diagnostics reveal characteristic power-law behaviour in the turbulent cascade, act as sensitive indicators of numerical instabilities undetected by conventional diagnostics, and expose localised structures in real space not apparent in the electrostatic potential alone. To support these analyses, we present a systematic comparison of explicit time-integration schemes, showing that the optimal choice depends critically on the regime: SSP22 achieves the lowest computational cost in linear ITG simulations, while SSP33 is most efficient in the nonlinear saturated state, with nonlinear simulations requiring two to three orders of magnitude more computational cost than linear cases. These results establish free-energy diagnostics as a valuable tool for interpreting plasma turbulence and provide practical guidance for scheme selection in \texttt{stella} simulations.
\end{abstract}

\section{Introduction}

In magnetic fusion plasma physics, one of the primary challenges for commercial reactor design is the reduction of turbulent transport~\cite{WhiteFNE2024,LionFED2025}. Microturbulence produces transport levels substantially larger than those predicted by classical collisional theories~\cite{HortonRMP1999,KrommesPR2002}, thereby degrading the confinement of heat, momentum, and particles and reducing the energy confinement time, \(\tau_E\). Microinstabilities can be driven linearly unstable by equilibrium density and temperature gradients~\cite{GarbetNF2010,RenPRL2011,JorgePPCF2021,ThienpondtNF2025}. The system can then saturate via nonlinear processes, such as the generation and shearing action of zonal flows~\cite{DiamondPPCF2005,PlunkNJP2017}, as well as the transfer of energy to damped modes~\cite{NavarroPRL2011,NavarroPoP2011}. A comprehensive description of these mechanisms is therefore crucial for understanding turbulent transport.

This turbulent behaviour emerges within a plasma characterised by a vast range of spatial and temporal scales. The macroscopic dimensions and transport scales of the device are orders of magnitude larger than the ion gyroradius and turbulent timescales. This allows us to separate different dynamics such as transport and turbulence. Within this framework, we can further average over the fast gyrophase and reduce the dimensionality of the phase space~\cite{GarbetNF2010,KrommesARFM2012}. This gyrokinetic framework will be outlined in \cref{sec:gke}.

While the theoretical framework of gyrokinetics is well established ~\cite{GarbetPPCP2004,RenRMPP2024,NaNRP2025}, turbulent transport remains a central challenge in magnetic-confinement devices, including both tokamaks and stellarators~\cite{HelanderPPCF2012,MynickPPCF2014,FujisawaPJA2021}. The inherent complexity of turbulent dynamics makes a complete analytical description currently impossible, necessitating the use of high-fidelity numerical simulations. However, these simulations remain computationally demanding as the system remains numerically stiff~\cite{SanchezNF2021}. In particular, the fast parallel streaming of electrons imposes a stringent timestep restriction on fully explicit schemes, motivating the use of implicit-explicit (IMEX) time-integration methods~\cite{BarnesJCP2019,MandellJPP2024}. 

In the IMEX approach, as applied to gyrokinetics, the stiff electron dynamics are treated implicitly, while the remaining terms are advanced explicitly. Consequently, for the numerical formulations used in this work, computational efficiency is strongly influenced by the stability and step-size constraints of the explicit integrator~\cite{BarnesJCP2019}. Implementing and evaluating robust explicit schemes is therefore an important step toward reducing the computational cost of simulations used in efficiently modelling plasma turbulence.

This paper addresses the challenges of modelling turbulence in plasmas through three primary goals: first, to implement and evaluate a variety of efficient explicit numerical algorithms within the \texttt{stella} code\cite{BarnesJCP2019}; second, to validate these algorithms using the Helmholtz free energy as a physics-based diagnostic of stability and energy conservation; and third, to use these tools to investigate mechanisms underlying turbulent transport.

The paper is organized as follows. \Cref{sec:gkm_and_ns} provides the theoretical foundation, covering the gyrokinetic framework, the role of free energy, and the fluxtube approximation. It further discusses the general principles of IMEX time-integration schemes before detailing the specific strategy implemented in the \texttt{stella} code. \Cref{sec:numerical_schemes} details the validation of the new numerical schemes through stability analysis and linear and nonlinear simulations. Finally, \Cref{sec:fe_diagnostic} demonstrates the capabilities of the Helmholtz free-energy diagnostic for a standard ion-temperature-gradient (ITG) turbulence case, providing a baseline for future investigations.

\section{Gyrokinetic Modelling and Numerical Setup\label{sec:gkm_and_ns}}
\subsection{The gyrokinetic equation\label{sec:gke}}

In the core of fusion-relevant plasmas, fluid descriptions may be insufficient because the dynamics involve the full particle distribution function rather than only a finite set of velocity moments. Kinetic effects such as wave-particle resonances, magnetic trapping, and finite-Larmor-radius physics can therefore play an important role, particularly in the low-collisionality regime. A kinetic treatment of the particle dynamics is consequently required
~\cite{BrizardRMP2007}. In kinetic theory, each plasma species is described by a distribution function \(f_{\nu} (\vec{x},\vec{v}, t)\), which evolves in the six-dimensional phase space \((\vec{x},\vec{v})\) according to the Boltzmann equation
\begin{equation}
     \frac{\mathrm{d} f_{\nu}(\vec{x},\vec{v}, t)}{\mathrm{d} t}  = \sum_{\nu'}C[f_{\nu},f_{\nu'}] \label{eq:boltzmann},
\end{equation}
where the left-hand side describes particle motion under electromagnetic forces, while the right-hand side represents collisions between species $\nu$ and $\nu'$~\cite{HelanderBook2005}. For notational simplicity, the time argument is suppressed in the following equations.

Directly solving \Cref{eq:boltzmann} is computationally prohibitive because magnetized plasmas exhibit a wide separation of spatial and temporal scales~\cite{AbelRPP2013,KrommesARFM2012}. Equilibrium quantities vary over macroscopic length scales like the minor radius \(a\), whereas turbulent fluctuations develop on the scale of the ion gyroradius \(\varrho_i = v_{\mathrm{th},i}/\Omega_{i}\), where \(v_{\mathrm{th},\nu} = \sqrt{2 T_{\nu} /m_{\nu}}\) is the thermal velocity, with \(T_\nu\) the temperature and \(m_\nu\) the mass of species \(\nu\), \(\Omega_{\nu}=Z_{\nu} e B_0/m_{\nu} c\) is the gyrofrequency, where \(Z_\nu\) is the charge state, \(e\) the elementary charge and \(c\) the speed of light, and \(i\) indicates the ion species. Likewise, the turbulent evolution occurs on timescales much longer than the rapid gyromotion \(\omega \ll \Omega_{\nu}\). These scale separations motivate an asymptotic expansion in the small parameter \(\rho_\ast  = \frac{\varrho_i}{a} \ll 1\).

To simplify the resulting equations, the total distribution function \(f_{\nu}\) of species \(\nu\) is decomposed into an equilibrium and a fluctuating contribution, also called the delta-f approach~\cite{DentonJCP1995}, as follows
\begin{equation}
    f_{\nu} = F_{\nu} + \delta \! f_{\nu} 
\end{equation}
where \(F_{\nu}\) is an equilibrium distribution~\cite{AbelRPP2013} and $\delta \! f_{\nu} $ is the fluctuation piece, which to leading order describes the turbulence. At leading order, the equilibrium is a local Maxwellian ,
\begin{equation}
F_{0,\nu}(\vec{x},v_\parallel,\mu)
=
\left[
\frac{m_\nu}{2\pi T_{0,\nu}(\vec{x})}
\right]^{3/2}
n_{0,\nu}(\vec{x})
\exp\left[
-\frac{
m_\nu v_\parallel^2/2+\mu B_0(\vec{x})
}{
T_{0,\nu}(\vec{x})
}
\right],
\end{equation}
where \(n_{0,\nu}\), \(T_{0,\nu}\), and \(B_0\) denote the equilibrium density, temperature, and magnetic-field strength, respectively. For notational simplicity, the arguments of equilibrium quantities are suppressed in subsequent expressions when they are clear from the context and it will be taken that these are flux functions that only depend on radial position throughout.

Here, we assume a static equilibrium magnetic field \(\vec{B}_0\) with magnitude \(B_0 = |\vec{B}_0|\) and unit vector \(\hat{\vec{b}} = \vec{B}_0/B_0\). The velocity is decomposed into a component parallel to the magnetic field, \(v_\parallel = \vec{v} \cdot \hat{\vec{b}}\), and a perpendicular component \(v_\perp\). The magnetic moment, defined as \(\mu = m_{\nu} v_\perp^2 / 2 B_0\), serves as a key adiabatic invariant in the magnetized limit.

This physical picture is encapsulated in the standard gyrokinetic ordering~\cite{FriemanPoF1982,KrommesARFM2012}
\begin{equation}
\frac{\delta \! f_{\nu}}{F_{0,\nu}} \sim \frac{Z_{\nu} e\delta \phi}{T_{0,\nu}} \sim \frac{k_\parallel}{k_\perp}
\sim \frac{\omega}{\Omega_{\nu}} \sim \varrho_* \ll 1, 
\label{eq:GK_ordering}
\end{equation}
where $\delta\phi$ is the fluctuating electrostatic potential, and $k_\parallel$ and $k_\perp$ denote the characteristic parallel and perpendicular wavenumbers.

Based on this ordering, the gyromotion occurs on a much faster timescale than the turbulent evolution~\cite{BrizardRMP2007}. The particle phase-space coordinates can therefore be transformed from \((\vec{x},\vec{v})\) to gyrocenter coordinates \((\vec{X},v_\parallel,\mu,\theta)\), where \(\theta\) is the gyroangle. Averaging over the fast gyroangle removes \(\theta\) as a dynamical coordinate, reducing the gyrokinetic phase space to \((\vec{X},v_\parallel,\mu)\) with \(\vec{X} = (\Psi, \alpha, z)\), where \(\Psi\) labels magnetic flux surfaces, \(\alpha\) is the field-line label, and \(z\) parametrises distance along the magnetic field. We then define \(g_\nu\)~\cite{BrizardRMP2007} by
\begin{equation}
    g_{\nu}(\vec{X},v_{\parallel},\mu) = \delta \! f_{\nu} + \frac{Z_{\nu} e}{T_{0,\nu}} \left[\delta\phi (\vec{x}) - \left\langle \delta\phi\right\rangle_{\vec{X}} (\vec{X}) \right] F_{0,\nu},
\end{equation}
where \(\delta f_\nu\) and \(F_{0,\nu}\) are understood to be expressed in gyrocenter coordinates and the gyroangle average at fixed gyrocenter position is 
\begin{equation}
  \langle A\rangle_{\vec{X}}  = \frac{1}{2\pi} \int_0^{2\pi} A(\vec{X}+\vec{\varrho}(\theta))\, \mathrm{d}\theta.
\end{equation}
Here, \(\vec{\varrho}(\theta)\) is defined as the gyroradius vector. The quantity \(g_\nu\) removes the adiabatic part of the perturbed distribution. The electrostatic gyrokinetic equation is then
\begin{equation}
\begin{split}
    &\frac{\partial g_{\nu}}{\partial t} + \underbrace{v_{\parallel}\hat{\vec{b}}\cdot \nabla z \left(\frac{\partial g_{\nu}}{\partial z} + \frac{Z_{\nu} e }{T_{0,\nu}} \frac{\partial\langle\delta\phi\rangle_{\vec{X}}}{\partial z} F_{0,\nu}\right)}_{\text{streaming}} - \underbrace{\frac{\mu}{m_{\nu}} \hat{\vec{b}} \cdot \nabla B_0 \frac{\partial g_{\nu}}{\partial v_{\parallel}}}_{\text{mirror}} \\
    & + \underbrace{\vec{v}_{ms} \cdot \left(\nabla_{\perp} g_{\nu} + \frac{Z_{\nu} e}{T_{0,\nu}} \nabla_{\perp} \langle\delta\phi\rangle_{\vec{X}} F_{0,\nu}\right)}_{\text{drifts}} + \underbrace{\langle\vec{v}_{E}\rangle_{\vec{X}} \nabla_{\perp} g_{\nu}}_{\text{nonlinearity}} + \underbrace{\langle\vec{v}_{E}\rangle_{\vec{X}} \nabla_{\perp} F_{0,\nu}}_{\text{drive}} \\
    =& \underbrace{-\mathcal{D}_{\mathrm{hyp}} g_\nu}_{\text{hyper-dissipation}},
\end{split}
\label{eq:gyrokinetic_in_g}
\end{equation} 
where the magnetic drift velocity $\vec{v}_{ms}$  is given by:
\begin{equation}
    \vec{v}_{ms} = \frac{c}{B_0} \hat{\vec{b}} \times \left( \frac{\mu}{m_{\nu}} \nabla B_0 + v_\parallel^2 \vec{\kappa} \right),
\end{equation}
and \(\vec{\kappa} =(\hat{\vec{b}}\cdot \nabla) \hat{\vec{b}}\) is the magnetic curvature. The fluctuating $\vec{E}\times\vec{B}$ drift velocity is:
\begin{equation}
    \vec{v}_E = \frac{c}{B_0} \hat{\vec{b}} \times \nabla_{\perp}\delta\phi.
\end{equation}

The electrostatic potential is determined self-consistently using the quasi neutrality condition
\begin{equation}
    \sum_{\nu} Z_{\nu} \delta n_{ \nu} = \sum_{\nu} Z_{\nu} \int\mathrm{d}^3 v \, \delta \! f_{\nu} = 0
\label{eq:quasineutrality}
\end{equation}
where the density fluctuations are defined as
\begin{equation}
    \delta n_{\nu} = \int \mathrm{d}^3 v \, \delta\!f_{\nu}. 
\end{equation}
The coupled system of the gyrokinetic \Cref{eq:gyrokinetic_in_g} and the quasi-neutrality condition \Cref{eq:quasineutrality} must be solved self-consistently to advance the distribution function~\cite{LeeJCP1987}.

\subsection{Free-energy balance and stability \label{sec:free_energy}}

The onset of gyrokinetic microinstabilities is fundamentally linked to the availability of free energy in equilibrium plasma gradients, which can be converted into fluctuation free energy\cite{HelanderJPP2022}. In the framework of gyrokinetics, this is quantified by the Helmholtz free energy \(H\), which is conserved for the undriven, collisionless gyrokinetic system~\cite{NavarroPRL2011,HelanderJPP2022}. Because this conservation law is independent of the magnetic geometry and the number of species, \(H\) provides a useful framework for analysing plasma stability and turbulence saturation~\cite{HelanderJPP2022,PlunkJPP2022}.

To formulate the free-energy balance, we introduce the non-adiabatic response \(h_{\nu}\)
\begin{equation}
    h_{\nu} =
    \delta \! f_{\nu}
    + \frac{Z_{\nu} e}{T_{0,\nu}}\delta\phi F_{0,\nu}
    =
    g_{\nu}
    + \frac{Z_{\nu} e}{T_{0,\nu}}
    \left\langle\delta\phi\right\rangle_{\vec{X}} F_{0,\nu}.
\end{equation}
Using this definition for \(h_{\nu}\), the Helmholtz free energy is defined as
\begin{equation}
    H =
    \Re\left[
    \sum_{\nu} T_{0,\nu}
    \left\langle
    \int
    \frac{ |h_{\nu}|^2 }{F_{0,\nu}}
    \,\mathrm d v_\parallel\,\mathrm d\mu\,
    2\pi\frac{B_0}{m_{\nu}}
    \right\rangle_\psi
    \right],\label{eq:free_energy}
\end{equation}
where \(\Re\) is the real part and
\begin{equation}
    \langle A\rangle_\psi =
    \frac{
    \displaystyle\int \mathrm{d} \alpha\, \mathrm{d} z
    \,\mathcal{J}(\alpha,z) A(\alpha,z)
    }{
    \displaystyle\int \mathrm{d} \alpha\, \mathrm{d} z
    \, \mathcal{J}(\alpha,z)
    },
\end{equation}
denotes the flux surface average with \(\mathcal{J}\) the Jacobian on the surface \(\psi=\mathrm{const.}\). The integration domains in \(\alpha\) and \(z\) are chosen to cover the complete flux surface.

The time evolution of \(H\) is obtained by multiplying the gyrokinetic \Cref{eq:gyrokinetic_in_g}, written in terms of \(h_{\nu}\), by the appropriate free-energy weight and integrating over phase space and the flux surface. The resulting balance decomposes the free-energy evolution into contributions from the equilibrium-gradient drive, conservative propagation and redistribution, and dissipation. In the present formulation, the drive term is
\[
\left\langle\vec v_E\right\rangle_\theta\cdot\nabla_\perp F_{0,\nu},
\]
where the gradient is taken at fixed \(v_\parallel\) and \(\mu\)
\begin{equation}
    \nabla_{\perp} F_{0,\nu}
    =
    F_{0,\nu}
    \left[
    \frac{\nabla_{\perp} n_{0,\nu}}{n_{0,\nu}}
    +
    \left(
    \frac{m_{\nu} v_\parallel^2}{2T_{0,\nu}}
    +
    \frac{\mu B_0}{T_{0,\nu}}
    -\frac{3}{2}
    \right)
    \frac{\nabla_{\perp}T_{0,\nu}}{T_{0,\nu}}
    \right].
\end{equation}

The equilibrium density and temperature gradients provide the thermodynamic free-energy source for microinstabilities. The resulting streaming, mirror, and magnetic-drift terms are conservative and redistribute free energy without making a net contribution to its global balance in the collisionless, undriven limit, provided the relevant boundary fluxes vanish. The streaming
term describes motion along the magnetic field, the mirror term arises from conservation of \(\mu\) in an inhomogeneous field, and the magnetic-drift terms describe magnetic-field-gradient and curvature effects~\cite{HelanderBook2005,Catto2019JPP}.

The nonlinear \(\vec E\times\vec B\) advection term redistributes free energy across perpendicular spatial scales and drives the turbulent cascade~\cite{SchekochihinPPCF2008}. In the analytical, driven collisionless system, the corresponding free-energy balance has the schematic form
\begin{equation}
\frac{\mathrm dH}{\mathrm dt}
=
\mathcal I_{\mathrm{drive}}
-
\mathcal D_{\mathrm{coll}},
\end{equation}
where boundary-flux terms are assumed to vanish.

In a numerical simulation, the turbulent cascade may transfer free energy to the smallest resolved scales, where it can accumulate near the spectral cutoff and produce spectral blocking ~\cite{JenkoNF2013,CrewsPoP2022}. The corresponding numerical free-energy balance can therefore be written schematically as
\begin{equation}
\frac{\mathrm dH}{\mathrm dt}
=
\mathcal I_{\mathrm{drive}}
-
\mathcal D_{\mathrm{coll}}
-
\mathcal D_{\mathrm{num}},
\label{eq:balance_num}
\end{equation}
where \(\mathcal D_{\mathrm{num}}\) denotes the net numerical contribution to the free-energy balance. This contribution may include both dissipation inherent in the numerical discretization and the effect of explicitly prescribed operators.

To control the accumulation of free energy near the spectral cutoff, a numerical hyper-dissipation term is introduced ~\cite{PueschelCPC2010,Hatch2011PRL} as
\begin{equation}
    \mathcal{D}_{\mathrm{hyp}} g_{\nu}
    =
    D_\perp\nabla_\perp^4 g_{\nu}
    +
    D_z\nabla_z^4 g_{\nu}
    +
    D_{v_\parallel}\nabla_{v_\parallel}^4 g_{\nu}.
\end{equation}
This operator is intended to provide a controlled sink at the smallest resolved scales, where physical collisionality is negligible.

Consequently, monitoring \(H\), together with the individual drive and dissipation contributions, provides a diagnostic of the stability and conservation properties of the numerical scheme. After accounting for the physical drive and the prescribed hyper-dissipation, a residual change in \(H\) can indicate errors associated with spatial discretization, boundary treatment, quasi-neutrality, or time integration~\cite{NavarroPoP2011,DoerkCPC2014}. The evolution of \(H\) can also provide information about nonlinear saturation and the redistribution of free energy among the excited modes~\cite{PlunkJPP2022}.

\subsection{Implicit-explicit (IMEX) time integration \label{sec:IMEX}}

Although the gyrokinetic approximation reduces the computational burden, solving the gyrokinetic equations remains computationally demanding because the gyrokinetic system still contains stiff terms. Efficient time-integration strategies are therefore required. Time-integration schemes for differential equations are generally classified into two broad categories: explicit and implicit methods~\cite{AscherANM1997}.

To assess the stability of a time-integration method, it is useful to consider the scalar Dahlquist test equation~\cite{DahlquistNumM1963}
\begin{equation}
\frac{\mathrm{d}u}{\mathrm{d}t}=\lambda u.    
\label{eq:dahlquist}
\end{equation}

The absolute-stability region of a method is defined as
\[
\mathcal{S}
=
\left\{
\zeta=\lambda\Delta t\in\mathbb{C}:
|R(\zeta)|\leq 1
\right\},
\]
where \(R(\zeta)\) is the stability function of the method. The region \(\mathcal{S}\) therefore specifies which combinations of time step and eigenvalue produce a stable numerical solution. This definition applies to both explicit and implicit methods.

Explicit schemes compute the state at the next timestep directly from the known state. For the test \Cref{eq:dahlquist} introduced above, the forward-Euler scheme reads 
\[
u^{n+1}
=
u^n+\Delta t\,\lambda u^n.
\]
They are computationally inexpensive per step because they do not require the solution of an algebraic system. However, the maximum stable timestep is determined by the intersection of the method’s stability region with the spectrum of the explicitly treated operator. For spatial discretizations of advective equations, this restriction is commonly expressed through a Courant-Friedrichs-Lewy (CFL) condition \cite{CourantMA1928}. Stiff terms can therefore force explicit schemes to use very small timesteps which can be computationally prohibitive. The stability regions of the explicit schemes considered in this work are discussed in detail in \Cref{sec:stability_regions}.

Implicit schemes, by contrast, evaluate the right-hand side using the unknown state at the new timestep. For the test \Cref{eq:dahlquist}, the backward-Euler method gives
\[
u^{n+1}
=
u^n+\Delta t\,\lambda u^{n+1}.
\]
This equation must be solved for \(u^{n+1}\), resulting in an algebraic system that is typically solved by using linear or nonlinear solvers. An implicit method is \(A\)-stable if its absolute stability region contains the closed left half-plane, \(\Re(\zeta)\leq 0\). An \(L\)-stable method is \(A\)-stable and additionally satisfies
\[
\lim_{\substack{|\zeta|\to\infty\\ \Re(\zeta)<\zeta}}
R(\zeta)=0.
\]
Thus, \(L\)-stable methods strongly damp highly stiff components rather than merely keeping them bounded~\cite{HairerBook1993}. Accuracy, nonlinear convergence, and the cost of the implicit solve may nevertheless impose practical timestep restrictions.

Given the advantages and limitations of both approaches, a common strategy for systems with disparate timescales is the use of mixed IMEX schemes~\cite{AscherANM1997,Kennedy2003ANM}. The semidiscrete evolution equation is decomposed schematically as
\[
\frac{\mathrm d u}{\mathrm dt}
=
\mathcal{F}(u)+\mathcal{G}(u),
\]
where \(\mathcal{F}\) contains the explicitly treated terms and \(\mathcal{G}\) contains the implicitly treated stiff terms. IMEX schemes therefore treat the stiff contribution implicitly to improve stability, while advancing the remaining contribution explicitly to avoid unnecessary computational cost. Both components are generally advanced using the same global timestep \(\Delta t\).

In gyrokinetic simulations, this distinction is particularly important because of the fast electron dynamics. The electron thermal speed is ordered \(v_{\mathrm{th},e}/v_{\mathrm{th},i}\sim\sqrt{m_i/m_e}\). For comparable electron and ion temperatures, this ratio is of order tens for hydrogenic ions and increases with ion mass. Consequently, electron parallel streaming occurs on a substantially faster timescale than ion dynamics and represents a significant source of stiffness \cite{BarnesJCP2019}. An explicit treatment of the corresponding terms would require time steps determined by the electron timescale, thereby increasing the computational cost. It is therefore advantageous to treat the stiff electron terms implicitly \cite{BarnesJCP2019,MandellJPP2024}. Other terms, such as the nonlinear \(\vec{E}\times\vec{B}\) advection, are commonly treated explicitly, although they may still impose a nonlinear CFL restriction.

For a fixed operator splitting and implicit method, the dominant cost of the implicit component is usually determined by the size and structure of the resulting algebraic systems. The explicit integrator can nevertheless strongly affect the overall efficiency through its stability region, number of stages, and allowable timestep. This motivates the primary objective of this paper: to investigate various explicit time-integration schemes and evaluate their potential to improve the computational performance of gyrokinetic simulations by permitting larger stable timesteps or providing more favourable stability properties for the explicit component. This will be addressed in \Cref{sec:numerical_schemes}

\subsection{The flux tube approximation}

The gyrokinetic ordering in \cref{eq:GK_ordering} establishes both the strong anisotropy of plasma turbulence, \(k_\parallel \ll k_\perp\), and the separation between turbulent and equilibrium scales. These scale separations motivate an elongated simulation domain along the magnetic field and a narrow domain in the perpendicular directions, forming the basis of the \enquote{flux tube} model for simulating plasma turbulence~\cite{Beer1995PoP,BallPPCF2021}.

The local domain is characterised by a radial extent \(\Delta x\), a binormal extent \(\Delta y\), and a parallel extent along the field line. The perpendicular dimensions are chosen to be small compared with the macroscopic equilibrium scale lengths, such as the density and temperature gradient lengths, defined by
\begin{equation}
    L_{n,\nu}
    =
    \frac{n_{0,\nu}}{|\nabla_\perp n_{0,\nu}|},
    \qquad
    L_{T,\nu}
    =
    \frac{T_{0,\nu}}{|\nabla_\perp T_{0,\nu}|}.
\end{equation}
At the same time, the domain must be sufficiently large to contain the dominant turbulent structures and several turbulent correlation lengths. Typical perpendicular box sizes can be of order \(100\rho_i\), whereas the perpendicular turbulent correlation length is typically of order several ion gyroradii, depending on the plasma parameters and turbulence regime~\cite{DiSienaNF2023}.

Within this local approximation, the equilibrium profiles are linearized about a specific flux surface. For example, the density and temperature profiles are represented locally as
\[
n_{0,\nu}(x)
\simeq
n_{0,\nu}(0)
+
x\left.\frac{\partial n_{0,\nu}}{\partial x}\right|_{x=0},
\qquad
T_{0,\nu}(x)
\simeq
T_{0,\nu}(0)
+
x\left.\frac{\partial T_{0,\nu}}{\partial x}\right|_{x=0}.
\]
Thus, the equilibrium gradients are retained as constant coefficients in the gyrokinetic equation \cref{eq:gyrokinetic_in_g}, while the equilibrium profiles are taken to be independent of \(y\) and \(z\) in the local approximation. The magnetic-geometry coefficients are evaluated on the flux surface but may vary along the field-line coordinate \(z\).

The fluctuating fields and distribution functions in the radial and binormal directions can then be treated using periodic boundary conditions. These boundary conditions are appropriate when the perpendicular domain is small compared with the equilibrium-gradient scale lengths, but sufficiently large to contain several turbulent correlation lengths~\cite{Beer1995PoP}.

Depending on the magnetic geometry and the chosen field-line domain, periodic or twist-and-shift boundary conditions may be imposed at the ends of the parallel coordinate. Periodic boundary conditions are appropriate when the parallel domain is geometrically periodic or when magnetic shear is neglected~\cite{XanthopoulosPRL2007,St-OngePPCF2023}. In toroidal configurations with magnetic shear, the twist-and-shift condition connects the two ends of the domain, accounting for global magnetic shear~\cite{Beer1995PoP}. When implemented consistently, both boundary conditions produce no net boundary contribution to the integrated free energy balance, since the associated free energy fluxes cancel. This local flux tube approach is widely used in gyrokinetic simulations because it retains the essential turbulent dynamics while avoiding the computational cost of a global calculation~\cite{BarnesJCP2019,MandellJPP2024}.

\subsection{Gyrokinetic code \texttt{stella}}

One code implementing an operator-split IMEX framework within a flux tube domain for \(\delta \! f_{\nu}\) gyrokinetics is \texttt{stella}~\cite{BarnesJCP2019}. The code employs a coordinate system and boundary conditions suitable for the flux tube geometry described above. The coordinate choice is \((x,y,z,v_\parallel,\mu)\), where \(z\) denotes the position along the magnetic field line, while \(x\) and \(y\) label the local radial and binormal directions, respectively.

The equilibrium is assumed to possess nested flux surfaces, with the label \(\Psi=\mathrm{const.}\) defining each flux surface~\cite{HelanderRPP2014}.
The equilibrium magnetic field is expressed using Clebsch coordinates as
\[
\vec{B} = \nabla \alpha \times \nabla \Psi,
\]
where \(\alpha\) is the field-line label~\cite{KruskalPoF1958}. The local perpendicular coordinates are linearly related to these labels according to
\begin{equation}
    x =
    \frac{\mathrm{d}x}{\mathrm{d}\Psi}
    \left(\Psi-\Psi_0\right),
    \qquad
    y =
    \frac{\mathrm{d}y}{\mathrm{d}\alpha}
    \left(\alpha-\alpha_0\right),
\end{equation}
where \(\Psi_0\) and \(\alpha_0\) denote the reference values at the center of the simulation domain. The precise definitions of \(x\) and \(y\) depend on the chosen magnetic-equilibrium configuration, and further details are provided in Barnes (2019)~\cite{BarnesJCP2019}.

\texttt{stella} represents the perpendicular dependence of the distribution function and fields using Fourier harmonics. This choice is well suited to statistically periodic perpendicular directions and allows us to retain spectral accuracy in the derivatives. The distribution function is expanded as
\begin{equation*}
g(x,y,z,v_\parallel,\mu,t)
=
\sum_{k_x,k_y}
\widetilde{g}_{\vec{k}}(z,v_\parallel,\mu,t)
\exp\!\left[
\mathrm{i}(k_xx+k_yy)
\right].
\end{equation*}

In velocity space, the distribution function is taken to vanish at the outer boundaries in \(v_\parallel\) and \(\mu\), consistent with the decay of the background Maxwellian. In the parallel direction, \texttt{stella} supports several boundary conditions, including periodic, twist-and-shift~\cite{Beer1995PoP}, and Martin boundary conditions~\cite{MartinPPCF2018}. The appropriate choice depends on the magnetic geometry and the parallel domain used in the simulation.

Within this framework, the gyrokinetic equation is decomposed into three parts to handle the disparate timescales efficiently. The first part contains the magnetic-drift terms, equilibrium-gradient drives, and nonlinear \(\vec{E}\times\vec{B}\) advection, and is treated explicitly. These terms are generally less stiff than the electron parallel streaming term. However each can still impose a CFL-type timestep restriction. The second part contains the mirror term and is advanced using a semi-Lagrangian method. This approach relaxes the CFL restriction associated with acceleration in \(v_\parallel\), although the timestep remains limited by accuracy and interpolation errors. The third part contains parallel streaming, which is treated implicitly because of the fast electron dynamics. This operator splitting combines the stability of an implicit treatment of the stiffest term with the efficiency of explicit treatment for the remaining terms~\cite{BarnesJCP2019}.

The operators are advanced sequentially using Lie splitting, with the output of each step providing the input to the next. All operators are advanced using the same global timestep \(\Delta t\)~\cite{McLachlanAN2002}. Standard Lie splitting is first-order accurate in \(\Delta t\). In \texttt{stella}, a \enquote{flip-flop} version is used in which the order of the operators is reversed on alternating timesteps~\cite{BoydBook2001,BarnesJCP2019}. For linear operators, this symmetric composition provides second-order accuracy over pairs of timesteps. When nonlinear operators are present, however, only first-order accuracy is generally guaranteed. This operator-splitting strategy is nevertheless effective for handling the disparate timescales in the gyrokinetic system.

The explicit advance in \texttt{stella} supports the standard fourth-order Runge--Kutta (RK4) method and second- and third-order strong-stability-preserving Runge--Kutta methods (SSPRK2 and SSPRK3) \cite{GottliebSIAM2001}.
\section{Time stepping schemes \label{sec:numerical_schemes}}

This section presents and evaluates different explicit integration schemes newly implemented in the gyrokinetic code \texttt{stella}. Because the IMEX algorithm introduced above separates the explicitly treated terms from the stiff components, the explicit integrators can be investigated independently. The schemes are evaluated using a reference case corresponding to an ITG instability.

The majority of the schemes considered here belong to the class of strong-stability-preserving (SSP) methods \cite{GottliebSIAM2001,CondeJSC2017}. These schemes are well suited to the nonlinear \(\vec{E}\times\vec{B}\) advection because their time step restriction can be related to the forward-Euler stability limit. If \(\Delta t_{\mathrm{FE}}\) denotes the maximum stable forward-Euler time step, an SSP method with coefficient \(\mathcal{C}\) is guaranteed to preserve the corresponding strong-stability property for
\[
\Delta t \leq \mathcal{C}\,\Delta t_{\mathrm{FE}}.
\] 
To evaluate the scheme with respect to computational cost per step, an effective SSP coefficient \(\mathcal{C}_{\mathrm{eff}}=\mathcal{C}/s\) is introduced, which accounts for the number of stages \(s\).

The eSSPRK\((2,2)\) and eSSPRK\((3,3)\) schemes (denoted SSP22 and SSP33) were
already established in \texttt{stella}. In this work, the
eSSPRK\({}^{+}(s,2)\) family for \(s \in [3,5]\) stages and the
eSSPRK\({}^{+}(4,3)\) scheme from \cite{IsherwoodSIAM2018} were
implemented. eSSPRK\({}^{+}\) schemes have non-decreasing abscissas,
which ensure monotonicity for all stages, but they have smaller SSP
coefficients than comparable eSSPRK schemes without this constraint.
Consequently, the newly implemented eSSPRK\({}^{+}\) schemes are not
expected to be more efficient than the existing eSSPRK schemes, but they
may offer advantages for certain nonlinear terms.

\subsection*{SSP schemes}

The explicit SSP Runge--Kutta schemes used here can be written in the Shu-Osher form
\begin{equation}
\begin{aligned}
    u^{(0)} &= u^{n}, \\
    u^{(i)} &= \alpha_i u^{(0)} + (1-\alpha_i)\left(
        u^{(i-1)} + \beta_i \Delta t\,F(u^{(i-1)})\right),
        \quad i=1,\dots,s, \\
    u^{n+1} &= u^{(s)},
\end{aligned}
\end{equation}
where \(F(u)\) is the semi-discretized right-hand side of the system of differential equations. For the eSSPRK\({}^{+}(s,2)\) family, the coefficients are
\[
\alpha_i = 0 \quad (i=1,\dots,s-1),\qquad
\beta_i = \frac{1}{s-1},
\]
and for the final stage
\[
\alpha_s = \frac{1}{s},\qquad
\beta_s = \frac{1}{s-1}.
\]
Thus the scheme reduces to
\begin{equation}
\begin{aligned}
    u^{(0)} &= u^{n}, \\
    u^{(i)} &= u^{(i-1)} + \frac{\Delta t}{s-1} F(u^{(i-1)}),
        \quad i=1,\dots,s-1, \\
    u^{(s)} &= \frac{1}{s} u^{(0)} + \frac{s-1}{s}\left(
        u^{(s-1)} + \frac{\Delta t}{s-1} F(u^{(s-1)})\right), \\
    u^{n+1} &= u^{(s)}.
\end{aligned}
\end{equation}
This family includes the eSSPRK\({}^{+}(3,2)\), eSSPRK\({}^{+}(4,2)\), and eSSPRK\({}^{+}(5,2)\) schemes. These schemes have the coefficient \(\mathcal{C}=s-1\).

The eSSPRK\({}^{+}(4,3)\) scheme is a third-order, four-stage method with coefficients given in \cite{IsherwoodSIAM2018}. In the same notation, it can be written as
\begin{equation}
\begin{aligned}
    u^{(0)} &= u^{n}, \\
    u^{(1)} &= u^{(0)} + \frac{11}{20}\Delta t\,F(u^{(0)}), \\
    u^{(2)} &= \frac{3}{8}u^{(0)} + \frac{5}{8}\left(
        u^{(1)} + \frac{11}{20}\Delta t\,F(u^{(1)})\right), \\
    u^{(3)} &= \frac{4}{9}u^{(0)} + \frac{5}{9}\left(
        u^{(2)} + \frac{11}{20}\Delta t\,F(u^{(2)})\right), \\
    u^{(4)} &= \frac{111}{1331}u^{(0)} + \frac{260}{1331}\left(
        u^{(0)} + \frac{11}{20}\Delta t\,F(u^{(0)})\right) \\
        &\quad + \frac{960}{1331}\left(
        u^{(3)} + \frac{11}{20}\Delta t\,F(u^{(3)})\right), \\
    u^{n+1} &= u^{(4)}.
\end{aligned}
\end{equation}
The SSP coefficients are summarized in \Cref{tab:ssp_efficiency}. 

\begin{table}[htbp]
    \centering
    \caption{SSP coefficient and effective SSP coefficient for the SSP schemes. The existing SSP22 and SSP33 schemes are eSSPRK variants, while the newly implemented SSP32, SSP42, SSP52, and SSP43 schemes are eSSPRK\({}^{+}\) variants.\label{tab:ssp_efficiency}}
    
    \begin{tabular*}{\columnwidth}{@{\extracolsep\fill} l c c c @{\extracolsep\fill}}
        \toprule
        \textbf{Scheme} & \textbf{Stages} & \textbf{\(\mathcal{C}\)} & \textbf{\(\mathcal{C}_{\mathrm{eff}}\)} \\
        \midrule
        SSP22 & 2 & 1 & 1/2 \\
        SSP33 & 3 & 1 & 1/3 \\
        SSP32 & 3 & 2 & 2/3 \\
        SSP42 & 4 & 3 & 3/4 \\
        SSP52 & 5 & 4 & 4/5 \\
        SSP43 & 4 & 20/11 & 5/11 \\
        \bottomrule
    \end{tabular*}
\end{table}

\subsection*{Non-SSP schemes}

In addition to the SSP schemes, the standard classical fourth-order Runge-Kutta method (RK4) is considered as a reference. We also consider the modified six-stage, fourth-order Runge-Kutta method (RK64) introduced for gyrokinetic applications by Doerk (2014)~\cite{DoerkCPC2014}.

The RK64 method can be written as
\begin{equation}
\begin{aligned}
    u^{(0)} &= u^{n}, \\
    u^{(1)} &= u^{(0)}, \\
    u^{(2)} &= u^{(0)} + 0.16791847\,\Delta t\,F(u^{(1)}), \\
    u^{(3)} &= u^{(0)} + 0.4829844\,\Delta t\,F(u^{(2)}), \\
    u^{(4)} &= u^{(0)} + 0.7054607\,\Delta t\,F(u^{(3)}), \\
    u^{(5)} &= u^{(0)} + 0.0929587\,\Delta t\,F(u^{(4)}), \\
    u^{(6)} &= u^{(0)} + 0.7621008\,\Delta t\,F(u^{(5)}), \\
    u^{n+1} &= u^{n} - 0.15108371\,\Delta t\,F(u^{(1)}) + 0.7538468\,\Delta t\,F(u^{(2)}) \\
    &\quad - 0.3601660\,\Delta t\,F(u^{(3)}) + 0.5269677\,\Delta t\,F(u^{(4)}) \\
    &\quad + 0.2304351\,\Delta t\,F(u^{(6)}).
\end{aligned}
\end{equation}
This method is designed to provide an improved stability region for problems with large imaginary eigenvalues, as encountered in hyperbolic and dispersive terms.

\subsection{Stability regions\label{sec:stability_regions}}

The stability regions introduced in \Cref{sec:IMEX} provide a useful framework for comparing these schemes. In particular, their extent along the imaginary axis is important for gyrokinetic applications, where weakly damped oscillatory modes can impose severe time step restrictions.

The linearized gyrokinetic equation, derived from  \Cref{eq:gyrokinetic_in_g}, can be formulated as an eigenvalue problem with a broad spectrum of complex eigenvalues. These eigenvalues characterize the timescales of the linear system. The nonlinear operator can also be associated with a characteristic timescale, as demonstrated for the gyrokinetic code \texttt{GENE} in~\cite{DoerkCPC2014}, building on the approach used in \cite{JenkoPoP2000}. Since collisionless gyrokinetic modes are often weakly damped, their frequencies can be much larger than their growth or damping rates. Consequently, a large stability-region extent along the imaginary axis is important for permitting useful time steps \cite{DoerkCPC2014}.

\begin{figure*}[htb]
    \centerline{\includegraphics[width=\linewidth]{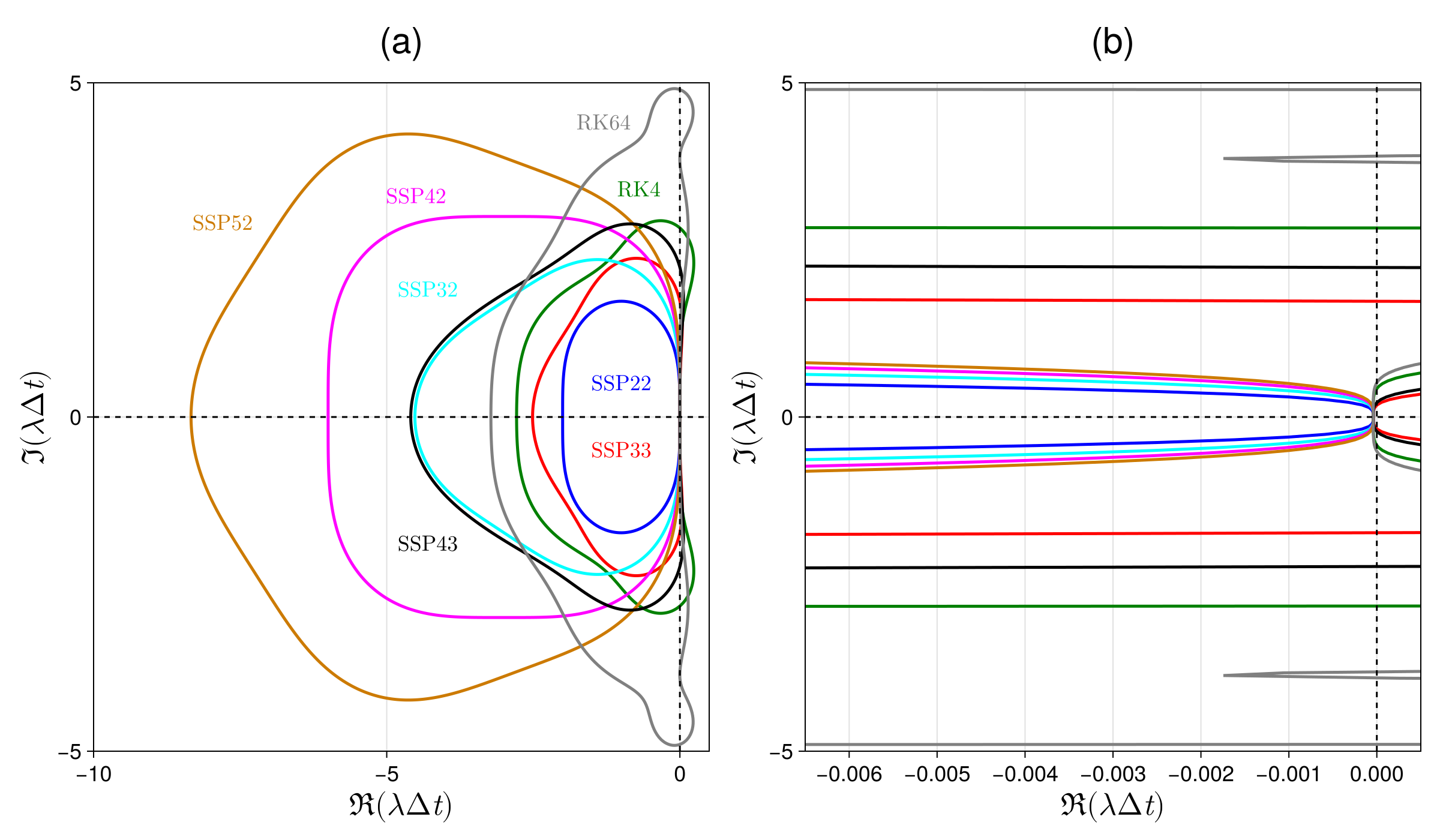}}
    \caption{Stability regions for explicit integration schemes plotted in the complex plane as \(\Re(\lambda \Delta t)\) vs. \(\Im(\lambda \Delta t)\), where \(\lambda\) is the eigenvalue of the test equation and \(\Delta t\) is the numerical time step. Panel (a) shows the overall regions and panel (b) shows a close-up near the origin. Each color corresponds to a different integration scheme.\label{fig:stability_region}}
\end{figure*}

\Cref{fig:stability_region} shows the stability regions for the different schemes in the complex plane. The axes represent the scaled eigenvalue \(z=\lambda\Delta t\) where \(\lambda\) is an eigenvalue of the linear operator and \(\Delta t\) is the numerical time step. The horizontal and vertical axes therefore correspond to \(\Re(\lambda\Delta t)\) and \(\Im(\lambda\Delta t)\), respectively. The real axis represents modal growth or damping, while the imaginary axis represents oscillatory dynamics. A time step is stable when all scaled eigenvalues of the discretized operator lie within the corresponding stability region.

The SSP schemes exhibit a large extent along the negative real axis, indicating their ability to damp modes with negative real eigenvalues. In comparison, RK64 shows a wider extent along the imaginary axis, making it a potentially stronger candidate for weakly damped oscillatory dynamics. Examining the zoomed region (b), one observes that all second-order SSP schemes are clustered around the real axis near the origin, while the third-order SSP schemes are broader in the imaginary direction near the origin. The fourth-order Runge-Kutta schemes, RK4 and RK64, have the largest extent near \(\Re(\lambda \Delta t)=-0.003\). The modified Runge--Kutta stability region also extends across the imaginary axis, which makes its stability boundary particularly sensitive to the location of weakly growing or weakly damped eigenvalues.

\Cref{tab:lin_efficiency} provides estimated values for the linear efficiency of the various schemes. The extent of the stability region along the imaginary direction is used as a proxy for the maximum stable time step for weakly damped modes. Here, this extent is evaluated at the intersection of the stability region with the line \(\Re(\lambda\Delta t)=-0.003\), corresponding to the region where many eigenvalues of the collisionless gyrokinetic operator were located in~\cite{DoerkCPC2014}.

For all schemes, the right-hand side must be evaluated once per stage. Since the additional operations associated with the intermediate Runge--Kutta updates are comparatively small, the computational cost is approximated as proportional to the number of stages. Under this assumption, the imaginary part of the intersection point divided by the number of stages provides a proxy for the linear performance. This linear efficiency metric is distinct from the effective SSP coefficient \(\mathcal{C}_{\mathrm{eff}}\) introduced earlier, which characterizes nonlinear strong-stability properties. The estimate is based on a fixed value of \(\Re(\lambda\Delta t)\) and is therefore intended as a relative comparison rather than an exact prediction of the maximum time step.

The results show that the second-order scheme (SSP22) and the second-order eSSPRK\({}^{+}\) family (SSP32, SSP42, SSP52) have low
linear efficiency and are thus stable only for small time steps when weakly damped oscillatory modes dominate. The best candidate appears to be the RK64 scheme mentioned earlier.

\begin{center}
\begin{table}[htbp] 
    \centering
    \caption{Intersection of the stability region with \(\Re(\lambda \Delta t) = -0.003\) as a performance indicator. The value corresponds to the region where most eigenvalues of the collisionless gyrokinetic equation were located in Doerk (2014)~\cite{DoerkCPC2014}. The efficiency proxy is defined as the imaginary part of the intersection point divided by the number of stages.\label{tab:lin_efficiency}}

    \begin{tabular*}{\columnwidth}{@{\extracolsep\fill} l S S S @{\extracolsep\fill}}
        \toprule
        \textbf{Scheme} & {\textbf{\(\Im(\lambda\Delta t)\)}} & {\textbf{Stages}} & {\textbf{Efficiency Proxy}} \\
        \midrule
        RK4   & 2.83 & 4 & 0.7075 \\
        SSP22 & 0.40 & 2 & 0.2    \\
        SSP33 & 1.74 & 3 & 0.58   \\
        SSP32 & 0.52 & 3 & 0.17   \\
        SSP42 & 0.60 & 4 & 0.15   \\
        SSP52 & 0.67 & 5 & 0.134  \\
        SSP43 & 2.25 & 4 & 0.5625 \\
        RK64  & 4.90 & 6 & 0.82   \\
        \bottomrule
    \end{tabular*}
\end{table}
\end{center}
It is important to note that the stability diagram is only valid for linear differential equations, so nonlinear stability properties may differ significantly from those discussed here. For SSP schemes, the strong-stability guarantee extends to nonlinear problems when the forward-Euler step satisfies an appropriate bound for the nonlinear operator~\cite{CondeJSC2017}. However, this guarantee applies to the relevant convex functional and time step restriction, rather than to the eigenvalue-based stability properties discussed in this section.

\subsection{Results for linear simulations}

To investigate the linear performance of the time-integration schemes, a W7-X case in the high-mirror vacuum configuration was prepared. The setup corresponds to a pure ion-temperature-gradient (ITG) mode with \(a/L_{T_i}=3\) and \(a/L_n=0\), a commonly studied case. Electrons were treated using the adiabatic modified Boltzmann response implemented in \texttt{stella}, thereby removing kinetic electron parallel dynamics and isolating the ITG mode. The flux tube was located at \(s=0.25\) and \(\alpha=0\) where \(s = \Psi/\Psi_{\mathrm{LCFS}}\) is the normalized toroidal flux and \(\Psi_{\mathrm{LCFS}}\) is the toroidal flux at the last closed flux surface.

The parallel coordinate was normalized using the minor radius \(a\), while the perpendicular turbulent coordinates were normalized using the reference thermal gyroradius, \(\varrho_{r} = v_{\mathrm{th},r}/\Omega_{r} = \sqrt{2T_{r}/m_{r}}/\Omega_r\). Time was normalized using the reference time \(t_r=a/v_{\mathrm{th},r}\), so that the dimensionless time step is defined as
\[
\Delta \hat{t}=\frac{\Delta t_{\mathrm{phys}}}{t_r}
=\Delta t_{\mathrm{phys}}\frac{v_{\mathrm{th},r}}{a}.
\]
All other normalisations can be found in the \texttt{stella} paper~\cite{BarnesJCP2019}.
The input parameters were set to \(n_{k_{y}} = 5\), \(n_{k_{x}} = 1\), \(k_{y,\mathrm{min}} = 1.0\), \(k_{y,\mathrm{max}} = 1.2\), \(n_z = 96\), \(n_{v_{\parallel}} = 60\), and \(n_{\mu} = 30\). Periodic boundary conditions were applied in the parallel direction, whereas the binormal and radial directions were treated as statistically periodic. The automatic time-step adaptation in the code, based on the CFL condition, was disabled to allow a direct assessment of the stability limitations of each scheme. Simulations were performed with time steps both below and above the SSP-based CFL estimate, as the SSP coefficient provides a sufficient but not necessary condition for stability. The simulations were terminated once one of the following criteria was satisfied:
\begin{itemize}
    \item The numerically estimated frequencies and growth rates of the relevant linear modes stabilized to a relative tolerance of \(10^{-4}\),
    \item A numerical instability caused the simulation to crash.\footnote{This occurs when the time step exceeds the stability limit of the explicit integrator, which is intentionally tested by disabling the automatic CFL-based time-step adaptation.}
\end{itemize}
Simulations that became unstable or crashed were excluded from the post-processing analysis.

The spectral benchmark was generated using the RK4 scheme with
\(\Delta\hat{t}=0.005\). Because the spectra represent time series data, they inherently contain some degree of statistical variability.
The comparison therefore used an inverse-variance-weighted root-mean-square error (RMSE) between the reference and test spectra. Specifically, each spectral point was assigned the weight \(w_i=1/\sigma_i^2\) where \(\sigma_i\) is the standard deviation of the corresponding test estimate. The weighted RMSE was therefore defined as
\[
\mathrm{RMSE}_{w}
=
\left[
\frac{\sum_i w_i
\left(y_i-y_i^{\mathrm{ref}}\right)^2}
{\sum_i w_i}
\right]^{1/2}.
\]
This weighting gives greater influence to spectral points with smaller statistical uncertainty. Simulations with an RMSE greater than \(10^{-4}\) were excluded from the analysis, as these exhibited significant deviations from the reference spectrum.

\begin{figure*}[htb]
    \centerline{\includegraphics[width=\linewidth]{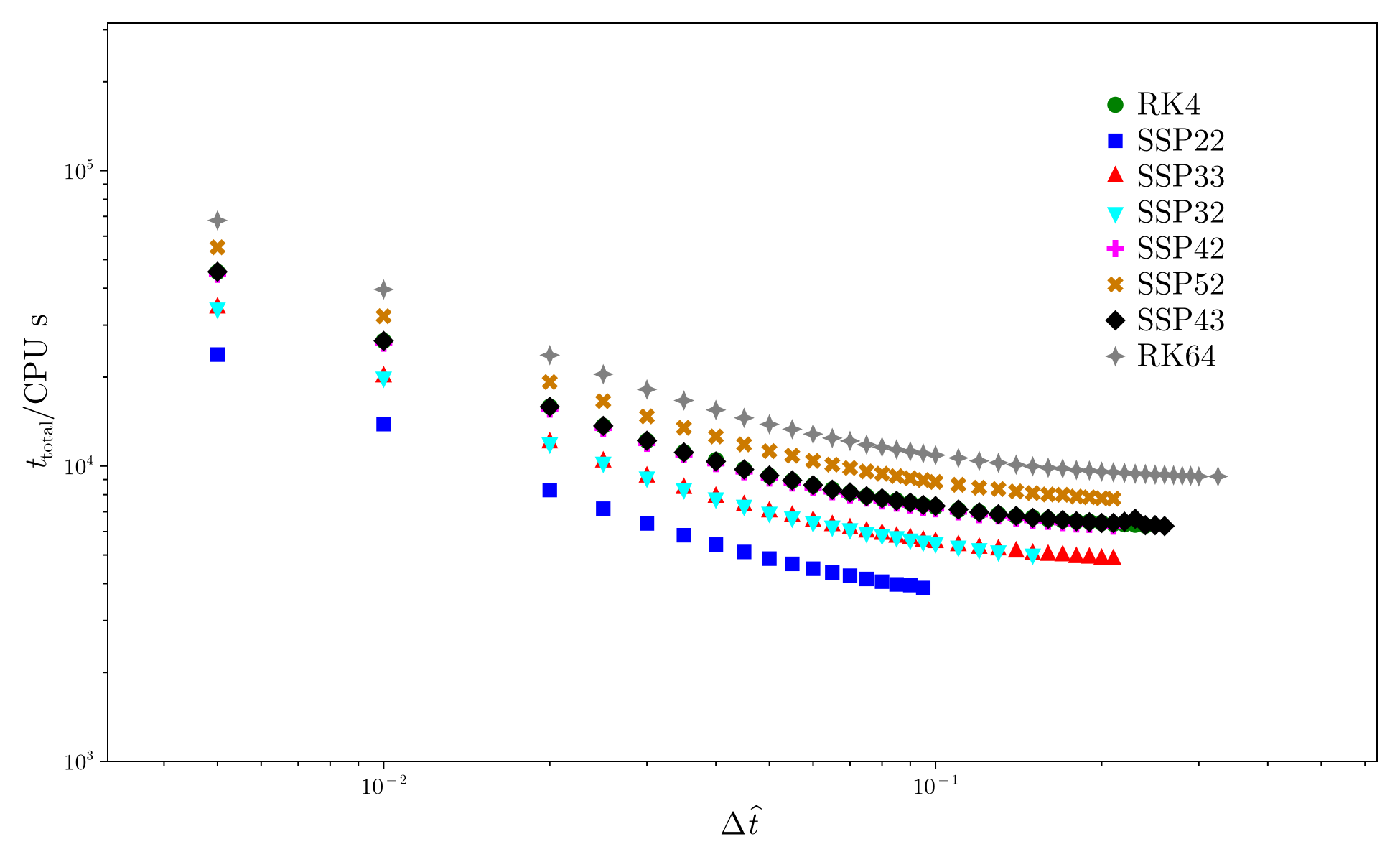}}
    \caption{Total simulation time \(t_{\mathrm{total}}\) as a function of the dimensionless time step \(\Delta \hat{t}\) for the different time integration schemes in the linear simulations. The colours are consistent with \Cref{fig:stability_region}.\label{fig:time_stel_lin}}
\end{figure*}

The total run times of the accepted simulations were then extracted and analysed as a function of the dimensionless time step \(\Delta \hat{t}\). The results are presented in \Cref{fig:time_stel_lin} on a log-log scale, and colours correspond to the schemes in \Cref{fig:stability_region}. Increasing \(\Delta \hat{t}\) reduces the total run time for each scheme until the largest stable time step tested is reached, because fewer steps are required to obtain stabilized frequency and growth rate spectra. At larger time steps beyond the stable range, the simulations fail before reaching the convergence criterion. This occurs because the time step exceeds the stability limit of the explicit integrator for the linear gyrokinetic operator, leading to exponential growth of numerical errors and eventual crash. The SSP coefficient provides a sufficient condition for stability, but the actual stability boundary depends on the spectrum of the discretized operator.

Schemes with the same number of stages often exhibit similar performance, despite having different stability limits. This observation is consistent with the assumption that the number of floating-point operations per time step remains approximately constant within a given stage count. For these linear simulations, SSP schemes with a larger number of stages do not demonstrate improved efficiency compared to schemes of similar order. The total run time \(t_{\mathrm{total}}\) scales approximately as \(t_{\mathrm{total}} \propto \Delta \hat{t} ^{-\xi}\), where the exponent \(\xi\) is approximately constant for each scheme within its stability region.

The performance of each scheme, characterised by the inverse of the total run time at its largest stable time step, is summarised in \Cref{tab:perf_data} together with the results of the nonlinear simulations.

The most computationally efficient scheme is SSP22, which achieves the shortest total run time over the range of stable time steps considered. This is notable because the second-order eSSPRK\({}^{+}\) schemes (SSP32, SSP42, SSP52) have substantially larger stability regions along the negative real axis than SSP22. The result highlights a trade-off: although a larger stability region permits larger time steps, the increased computational cost per step can offset this advantage. For this case, schemes with fewer stages are therefore more computationally efficient. Furthermore, the run times of RK64 and SSP52 appear to saturate at large time steps, suggesting that the total simulation time becomes dominated by fixed overhead associated with the convergence check rather than by the number of time steps.

The observed efficiency ranking can be understood in terms of the
interplay between the stability region and the computational cost per step. SSP22 has the smallest number of stages (two) and therefore the lowest cost per step, but its stability region along the imaginary axis is relatively small. The eSSPRK\({}^{+}(s,2)\) family has larger stability regions along the negative real axis, which would permit larger time steps for strongly damped modes, but the additional stages increase the computational cost. In the present linear ITG case, the eigenvalues are predominantly weakly damped and lie close to the imaginary axis, as discussed in \Cref{sec:stability_regions}. Consequently, the advantage of a larger real-axis stability region is not fully utilised, and the lower stage count of SSP22 becomes the dominant factor. RK64, despite having the largest stability region along the imaginary axis, requires six stages and therefore has a higher computational cost per step. In the linear regime, this additional cost outweighs the benefit of its larger stability region, resulting in lower efficiency than SSP22.

\subsection{Results for nonlinear simulations}\label{sec:nonlinear_sim}
The nonlinear simulations considered the same ITG case and flux tube as the linear simulations. The numerical parameters were set to \(n_{y} = 74\), \(n_{x} = 60\), \(k_{y,\mathrm{min}} = 0.05\), \(n_z = 96\), \(n_{v_{\parallel}} = 60\), and \(n_{\mu} = 30\). Periodic boundary conditions were applied in the parallel direction. The automatic time-step adaptation in the code based on the CFL condition was disabled to allow a direct assessment of the stability limitations of each scheme. 

An RK4 simulation with \(\Delta \hat{t}=0.05\) was first evolved into the nonlinear quasi-steady saturated phase, reaching approximately \(\hat{t}=500\). A checkpoint from this phase was then used to initialise simulations with the different time-integration schemes and time steps. Each restarted simulation was evolved for the same normalized time interval, \(\Delta\hat{t}_{\mathrm{run}}=400\), rather than for the same number of time steps. During these runs, the flux tube integrated free energy, \(H\), and its time derivative, \(\mathrm{d}H/\mathrm{d}t\), were computed.

Using a common saturated-state checkpoint ensured that differences between the simulations were primarily associated with the time-integration scheme rather than with differences in their preceding transient evolution.

The original RK4 simulation with \(\Delta\hat{t}=0.05\) was used as the reference case. The time-averaged free energy of each test simulation was compared with that of the reference. Simulations for which
\[
\delta_H =
\frac{\left|\langle H\rangle_{t}-\langle H_{\mathrm{ref}}\rangle_{t}\right|}
{\langle H_{\mathrm{ref}}\rangle_{t}}
>0.2
\]
were excluded from the analysis. Here, the averages were evaluated over the corresponding analysis intervals. This ensured that only systems in the same physical state were compared. The temporal evolution of the free energy was also examined to verify that the simulations remained in a statistically stationary saturated state.

\begin{figure*}[htb]
    \centerline{\includegraphics[width=\linewidth]{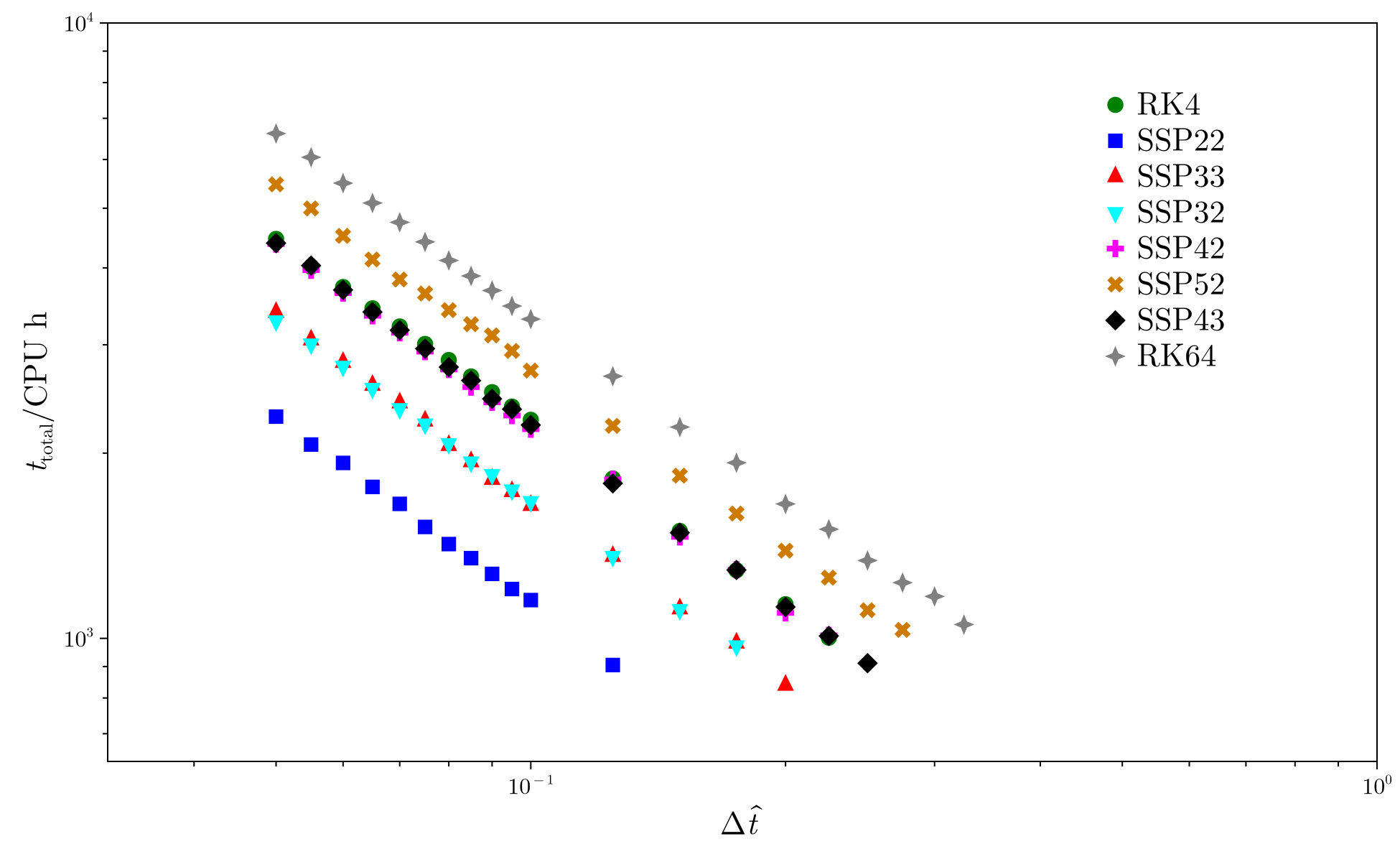}}
    \caption{Total simulation time \(t_{\mathrm{total}}\) as a function of the dimensionless time step \(\Delta \hat{t}\) for the different time integration schemes in the nonlinear simulations. The colours are consistent with \Cref{fig:stability_region}.\label{fig:time_stel_non}}
\end{figure*}

The runtimes of the accepted nonlinear simulations were then evaluated as a function of the dimensionless time step \(\Delta \hat{t}\), as presented in \Cref{fig:time_stel_non}. The results are presented on a double-logarithmic scale, with colours indicating the corresponding schemes from \Cref{fig:stability_region}. As in the linear case, increasing \(\Delta \hat{t}\) generally reduced the total runtime because fewer time steps were required to complete the prescribed simulation interval. However, the nonlinear simulations have substantially longer runtimes due to the increased computational cost of evaluating the nonlinear terms.

The most computationally efficient scheme is SSP33, which achieves the shortest runtime over the range of stable time steps considered. SSP22 is the second most efficient scheme and performs particularly well at small time steps, but fails for \(\Delta\hat{t}\geq 0.15\), consistent with the stability behaviour observed in the linear simulations. RK4 and SSP32 exhibit similar performance. As in the linear simulations, the higher-order schemes do not provide a clear efficiency advantage in this nonlinear case.

The change in the most efficient scheme from SSP22 in the linear case to SSP33 in the nonlinear case can be understood in terms of the different stability constraints in the two regimes. In the linear ITG case, the eigenvalues are predominantly weakly damped and lie close to the imaginary axis, favouring schemes with low stage counts. In the nonlinear saturated regime, the time step restriction is determined not only by the linear eigenvalues but also by the nonlinear \(\vec{E}\times\vec{B}\) advection term. This nonlinear term introduces additional stability constraints that depend on the amplitude of the turbulent fluctuations and the strength of the zonal flows. SSP33, despite having three stages, has a larger stability region along the imaginary axis than SSP22, which allows it to take larger time steps in the presence of the nonlinear terms. The eSSPRK\({}^{+}(4,3)\) scheme (SSP43) also performs well in the nonlinear case, achieving the third-best efficiency despite its relatively small effective SSP coefficient. This suggests that the stability region shape, particularly its extent along the imaginary axis, is more important than the SSP coefficient for this nonlinear problem. The results indicate that the optimal scheme depends on the specific balance between linear and nonlinear dynamics in the system. 

A comprehensive summary of the performance metrics for all tested time-integration schemes is provided in \Cref{tab:perf_data}. This table lists the largest tested stable time step, \(\Delta \hat{t}_{max}\), and the corresponding computational runtime for both the linear and nonlinear regimes.

\begin{table*}[!ht]
    \centering
    \caption{Summary of the best performance, defined as the shortest total simulation time, and the largest tested stable dimensionless time step \(\Delta \hat{t}\) for each time integration scheme in the linear and nonlinear simulations. Linear runtimes are given in CPU seconds, and nonlinear runtimes are given in CPU hours.\label{tab:perf_data}}
    
    \begin{tabular*}{\textwidth}{@{\extracolsep\fill} l S[table-format=1.3] S[table-format=4.0] S[table-format=1.3] S[table-format=4.0] @{\extracolsep\fill}}
        \toprule
        \textbf{Scheme} & {\textbf{\(\Delta \hat{t}_{\mathrm{max},\mathrm{linear}}\)}} & {\textbf{Linear runtime (CPU s)}} & {\textbf{\(\Delta \hat{t}_{\mathrm{max},\mathrm{nonlinear}}\)}} & {\textbf{Nonlinear runtime (CPU h)}} \\
        \midrule
        RK4   & 0.240 & 6268 & 0.225 & 1002 \\
        SSP22 & 0.095 & 3866 & 0.125 & 905  \\
        SSP33 & 0.210 & 4816 & 0.200 & 839  \\
        SSP32 & 0.150 & 5032 & 0.175 & 972  \\
        SSP42 & 0.210 & 6264 & 0.225 & 1011 \\
        SSP52 & 0.210 & 7754 & 0.275 & 1032 \\
        SSP43 & 0.260 & 6264 & 0.250 & 911  \\
        RK64  & 0.325 & 9223 & 0.325 & 1053 \\
        \bottomrule
    \end{tabular*}
\end{table*}

The efficiency rankings differ between the two cases. For the linear simulations, the schemes are ordered from shortest to longest runtime as SSP22, SSP33, SSP32, SSP42, SSP43, RK4, SSP52, and RK64. SSP22 is therefore the most computationally efficient scheme in the simulated linear regime.

For the nonlinear simulations, the corresponding ordering is SSP33, SSP22, SSP43, SSP32, RK4, SSP42, SSP52, and RK64. SSP33 is therefore the most efficient scheme in the nonlinear case. This indicates that the balance between computational cost per step and the allowable time step depends on whether the system is evolved in the linear or nonlinear saturated regime.

Although the largest tested stable time steps are broadly comparable between the two regimes, the nonlinear simulations require substantially more computational time. The linear runtimes range from approximately 4000 to 9000 CPU seconds, while the nonlinear runtimes range from approximately 840 to 1050 CPU hours, representing an increase of roughly two to three orders of magnitude. This reflects the additional cost of evaluating and advancing the nonlinear terms, as well as the larger number of cores required for the nonlinear simulations. The change in the most efficient scheme from SSP22 in the linear case to SSP33 in the nonlinear case further indicates that, in the nonlinear saturated regime, the larger stable time step of SSP33 compensates for its additional per-step cost.

\section{Free-energy diagnostic\label{sec:fe_diagnostic}}
The free-energy diagnostic implemented in \texttt{stella}, based on \Cref{eq:free_energy}, is used to examine three complementary aspects of gyrokinetic simulations. First, its spectral representation is compared with the expected power law behaviour of the turbulent cascade, providing a basic benchmark for the implementation. Second, the diagnostic is used to identify numerical instabilities that are not necessarily apparent in conventional diagnostics. Finally, the spatial and spectral distribution of free energy is used to provide an preliminary interpretation of the mechanisms involved in turbulent transfer.

Standard diagnostics, such as the electrostatic-potential amplitude, growth rate, and heat flux, quantify observable properties of the simulated turbulent state. For example, fluctuation amplitudes and spectra describe the spatial and scale distribution of the fields, whereas heat flux quantifies the associated radial transport. These quantities are essential for characterising gyrokinetic simulations, but they do not by themselves resolve the energetic pathways that maintain or destabilise the turbulent state.

The free-energy diagnostic provides a complementary, physics-based framework. Free energy is injected by equilibrium gradients and removed by dissipative processes, while the nonlinear term redistributes it among modes and scales without changing the global free energy. Thus, the free-energy balance links observed fluctuation and transport levels to the competing effects of drive, nonlinear redistribution, and dissipation. It also provides a sensitive check of numerical behaviour: departures from the expected balance can reveal numerical instabilities or excessive numerical dissipation that may not be immediately apparent in conventional observables. The formal definition and balance of the free energy were introduced in \Cref{sec:free_energy}. Here, we focus on their use as diagnostics of numerical behaviour and turbulent dynamics. \Cref{tab:diagnostics_comparison} summarises the complementary roles of standard and free-energy-based diagnostics considered in this work.

\begin{table*}[!ht]
    \centering
    \caption{Complementary roles of standard and free-energy-based diagnostics in gyrokinetic simulations. Standard diagnostics characterise field amplitudes and transport, whereas free-energy diagnostics assess the associated energetic content, balance, and nonlinear redistribution.\label{tab:diagnostics_comparison}}
    \begin{tabular*}{\textwidth}{@{\extracolsep\fill}
        p{0.16\textwidth}
        p{0.37\textwidth}
        p{0.37\textwidth}
        @{\extracolsep\fill}}
        \toprule
        \textbf{Aspect} &
        \textbf{Standard diagnostics} &
        \textbf{Free-energy diagnostics} \\
        \midrule
        Fluctuation level & Quantities such as \( |\phi|^2 \) characterise the amplitude and spatial structure of electrostatic fluctuations. & Free energy includes both field and distribution-function contributions, thereby characterising the energetic content of
        the perturbations. \\
        \\
        Transport & Heat and particle fluxes quantify the net radial transport produced by the turbulence. & The free-energy balance relates the turbulent state to energetic drive, nonlinear redistribution, and dissipation. \\
        \\
        Scale dependence & Field spectra identify the scales containing substantial fluctuation power. & Free-energy spectra and transfer terms identify how free energy is distributed and redistributed across scales. \\
        \\
        Numerical behaviour & Time traces, growth rates, and saturation levels reveal obvious instabilities or convergence problems. & Conservation and balance properties provide a physics-based test for numerical instabilities, unresolved transfer, or excessive numerical dissipation. \\
        \bottomrule
    \end{tabular*}
\end{table*}

The following subsections demonstrate these applications. We first use the free-energy spectrum as a basic benchmark of the turbulent cascade. We then examine its sensitivity to numerical instabilities and conclude with examples that illustrate how free-energy transfer can help interpret turbulent dynamics.

\subsection{Spectral benchmark of the free energy diagnostic}

As a basic benchmark of the newly implemented diagnostic, we examine the spectral structure of the free energy in a nonlinear gyrokinetic simulation. This test is intended to provide confidence that the diagnostic correctly captures the transfer of free energy across perpendicular spatial scales. It is not intended as a precise determination of universal spectral exponents.

The free energy is formally defined as an average over the full flux surface of the squared non-adiabatic response \(h_{\nu}\), see \Cref{eq:free_energy}. In the flux-tube simulations, it was approximated by the corresponding integral over the simulated flux tube
\begin{equation}
    \langle A\rangle_{\psi} \approx \langle A \rangle_{\text{FT}},
\end{equation}
where 
\begin{equation}
    \langle A\rangle_{\text{FT}} = \frac{\int \mathrm{d}k_x\, \mathrm{d}k_y\, \mathrm{d}z \, \mathcal{J}(k_x,k_y,z) A(k_x,k_y,z)  }{\int \mathrm{d}k_x\, \mathrm{d}k_y\, \mathrm{d}z \, \mathcal{J}(k_x,k_y,z) }
\end{equation}
denotes the flux tube average. This approximation is appropriate here because the integrated free-energy balance has no net boundary contribution: the parallel boundary flux is assumed to vanish, and the statistically periodic binormal and radial directions provide no net contribution after integration.

In turbulent gyrokinetic systems, equilibrium gradients inject free energy at relatively large scales, while nonlinear interactions redistribute it toward smaller spatial and velocity-space scales. The free energy cascades until it reaches scales at which physical
or numerical dissipation becomes effective. In an appropriate inertial or kinetic range, this transfer can produce characteristic power-law spectra~\cite{SchekochihinTAJSS2009,NavarroPRL2011,PlunkPRL2017}.

The free-energy spectrum, as well as the spectra of its constituent field and distribution-function contributions, can exhibit power-law behaviour across suitable wavenumber ranges. In particular, the ion-gyroradius scale separates different kinetic regimes, so that the spectral slope can change when \(k_{\perp}\varrho_r\) becomes of order unity. The observation of a consistent spectral decay and a change in slope near this scale provides a qualitative benchmark for the implementation of the free-energy diagnostic.

\begin{figure*}[htb]
    \centerline{\includegraphics[width=\linewidth]{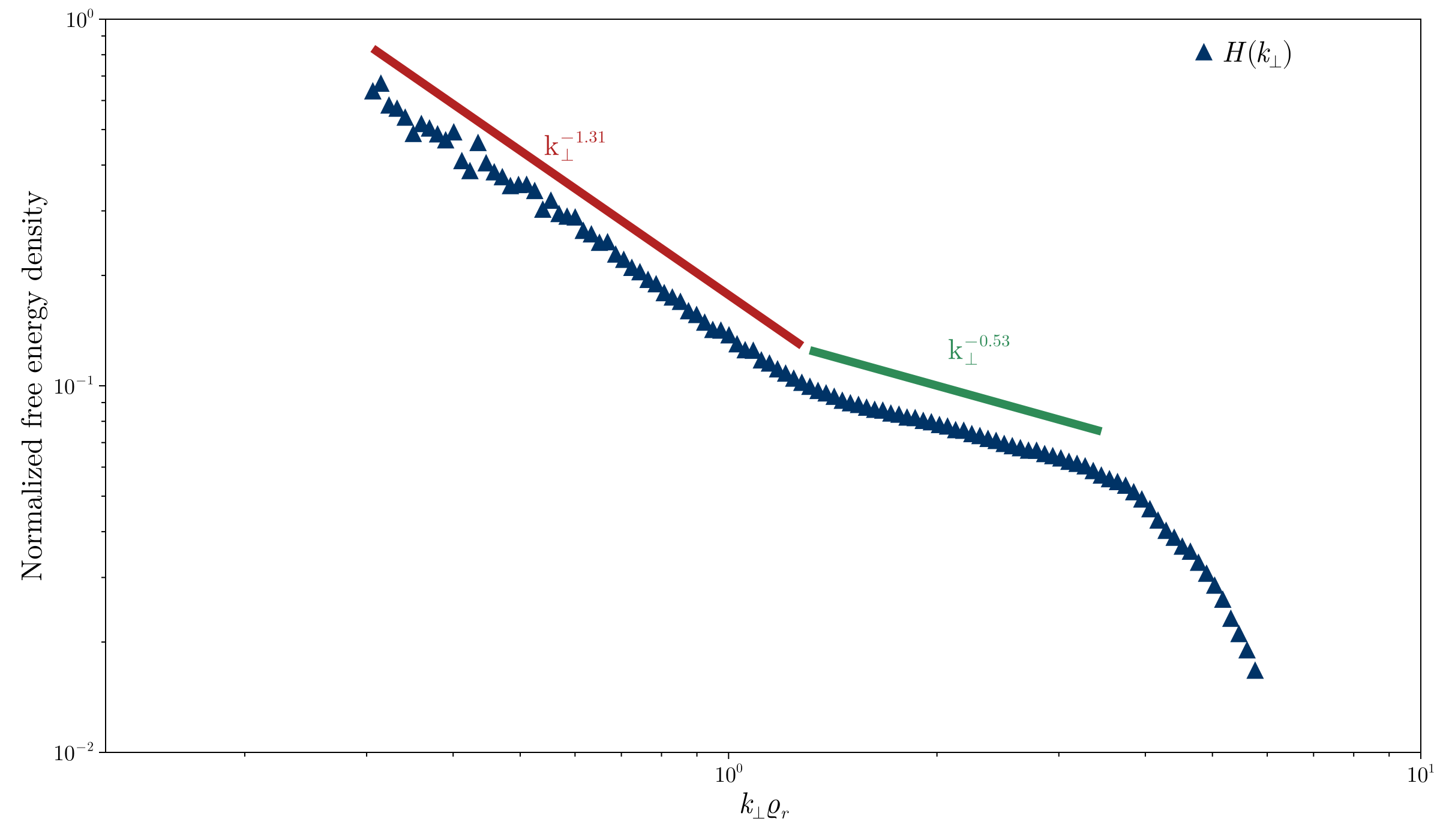}}
    \caption{Free-energy spectrum as a function of the perpendicular wavenumber \(k_{\perp}\). The blue triangles represent the numerical data for the high-mirror W7-X configuration, while the red and green lines are guides to the eye indicating approximate power-law behaviour over two wavenumber ranges.\label{fig:powerlaw}}
\end{figure*}

An example obtained with \texttt{stella} is shown in \Cref{fig:powerlaw} for \(a/L_{T_{i}}=3.5\) and \(a/L_{n}=1.0\) for the same high-mirror W7-X configuration introduced before. The free-energy spectrum is plotted as a function of the perpendicular wavenumber \(k_{\perp}\). The blue triangles represent the numerical data, while the red and green lines show power-law fits over two separate wavenumber ranges. The purpose of these fits is to illustrate the qualitative spectral structure; a detailed determination of the exponents and their dependence on the simulation parameters is beyond the scope of this work.

The spectrum exhibits an overall decay across the kinetic range, with three distinguishable regimes. A strong decay is observed for approximately \(0.3<k_{\perp}<1.3\), followed by a slower decline for \(1.3<k_{\perp}<3.5\). After that, an even faster decay sets in. We identify the first regime with the free-energy cascade found in \cite{NavarroPRL2011}, since the exponents are consistent. The second regime can be associated with the onset of finite Larmor radius effects. The last steep decay is caused by the perpendicular hyper-dissipation operator.

The observed spectral decay and the change in slope provide the expected qualitative signature of a gyrokinetic free-energy cascade. This agreement provides a basic consistency check that the implemented diagnostic captures scale-dependent free-energy transfer.
\subsection{Free energy as numerical stability diagnostic}

A significant advantage of the free-energy diagnostic is its high sensitivity to numerical instabilities, which may remain undetected when monitoring standard outputs such as turbulent heat fluxes and moments of the distribution function. To demonstrate this, we performed a nonlinear simulation based on the Cyclone Base Case~\cite{DimitsPoP2000}, similar to the setup described in \Cref{sec:nonlinear_sim}. Once the simulation reached a saturated state at \(t=500\), we deliberately altered the integration scheme (using SSP53) and the time step (\(\Delta t = 0.005\)). In a numerically stable configuration, such changes to the integration parameters should not significantly perturb the saturated values of the physical quantities.

\begin{figure*}[htb]
    \centerline{\includegraphics[width=\linewidth]{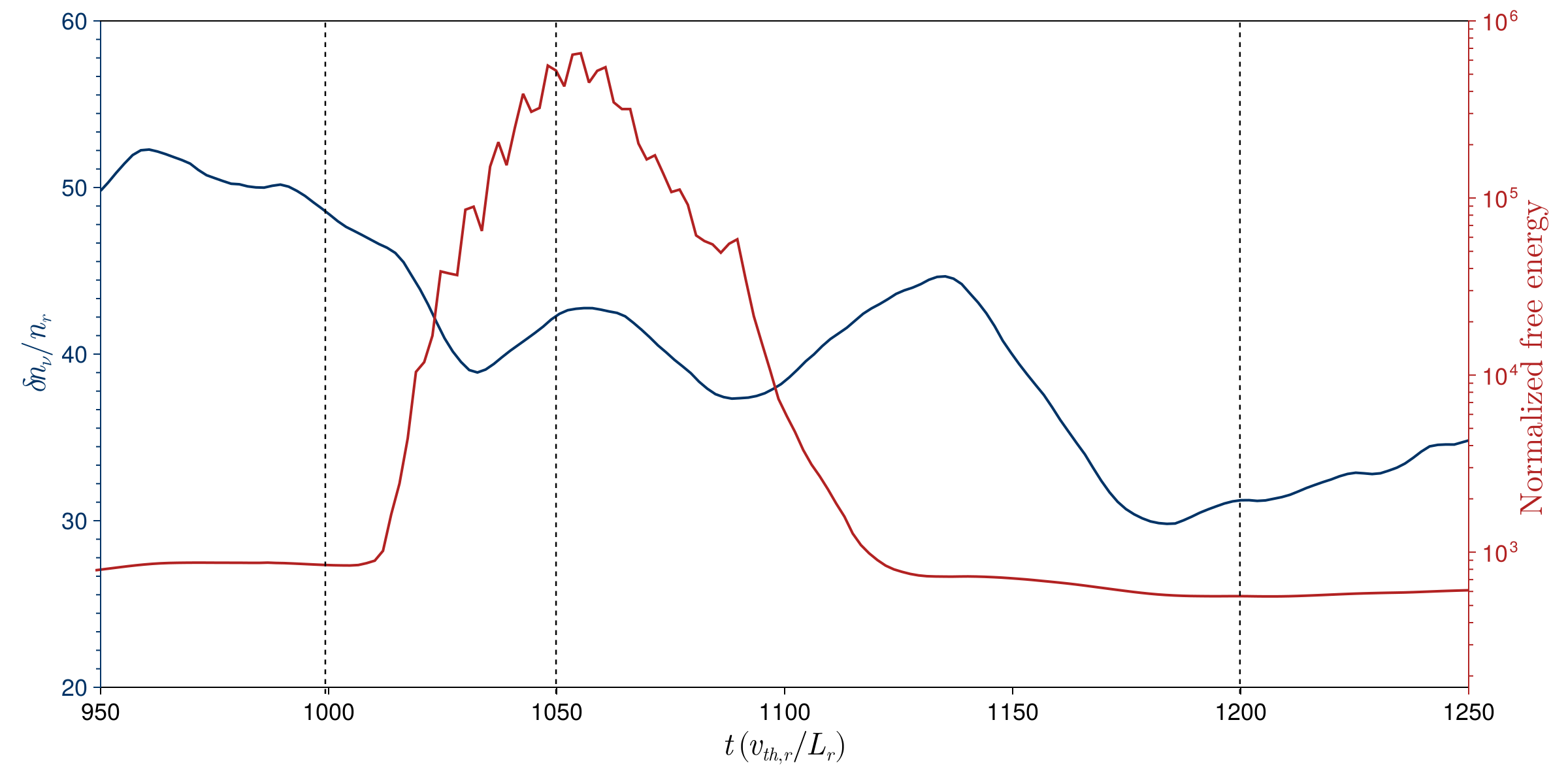}}
    \caption{Time evolution of the normalised density fluctuations (blue) and the normalised total free energy (red) for the Cyclone Base Case~\cite{DimitsPoP2000} with normalized gradients \(a/L_{T_i}=3.0\) and \(a/L_{n_i}=0\).\label{fig:num_inst_t}}
\end{figure*}

The resulting time trace in \Cref{fig:num_inst_t} reveals a strong difference between the density fluctuations and the free energy. While the density fluctuations, plotted on a linear scale, remain relatively smooth and suggest a stable saturated state, the free energy, plotted on a logarithmic scale, exhibits pronounced spikes. Even with the compressive nature of the logarithmic axis, these spikes are clearly visible, indicating a numerical instability that is completely invisible in the standard diagnostic. They are also absent in other conventional diagnostics, like the other moments of the distribution function, electromagnetic fields and fluxes. The vertical dashed lines mark the specific time snapshots that are analysed in the subsequent figures.

\begin{figure*}[htb]
    \centerline{\includegraphics[width=\linewidth]{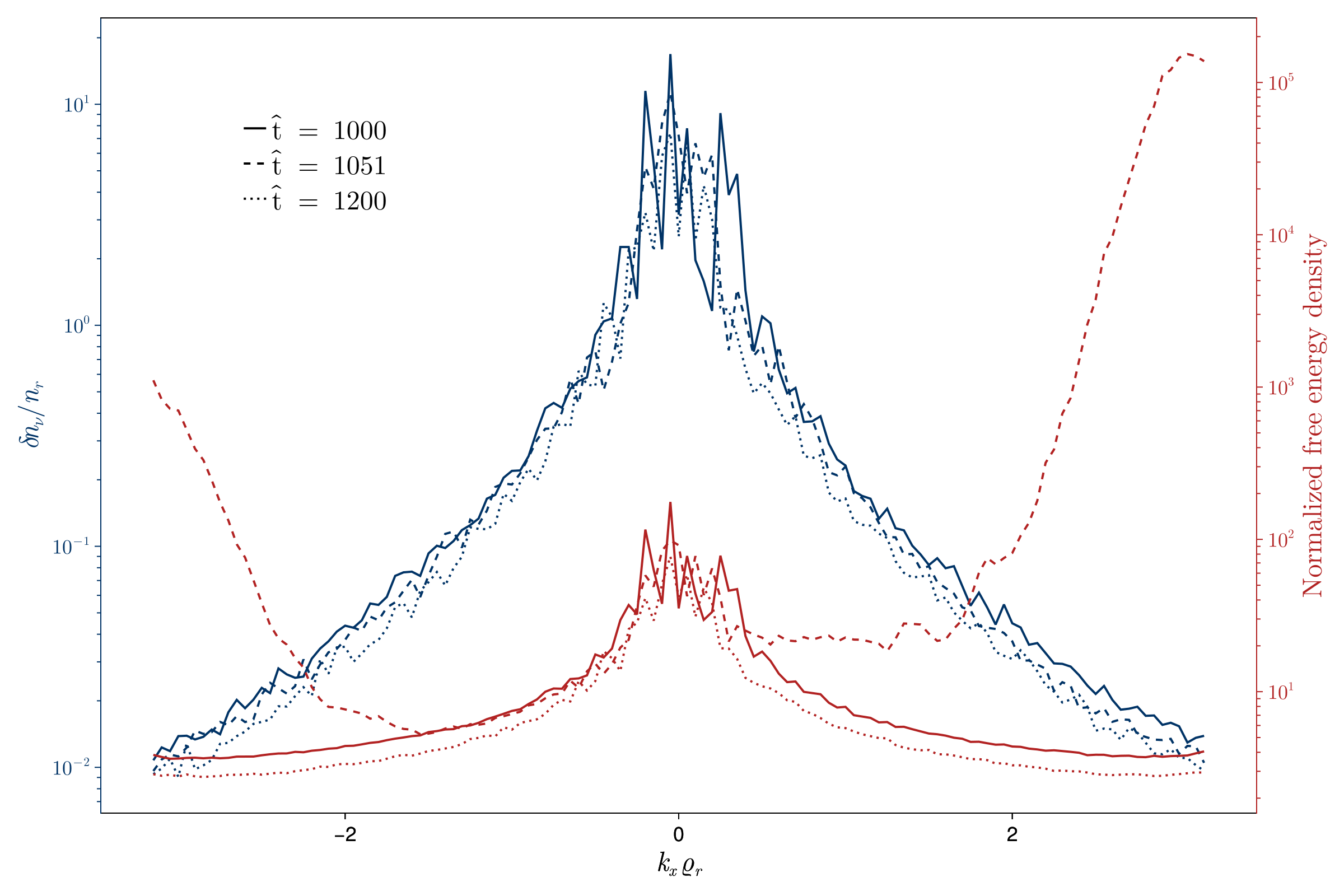}}
    \caption{Spectral distribution of the normalised density fluctuations (blue) and the free energy (red) along the radial direction at three different time steps, integrated over velocity space and the parallel and binormal directions: \(t = 1000\) (prior to the onset of the instability), \(t=1051\) (during the spike), and \(t=1200\) (following the instability).\label{fig:num_inst_kxkyz}}
\end{figure*}

\begin{figure*}[htb]
    \centerline{\includegraphics[width=\linewidth]{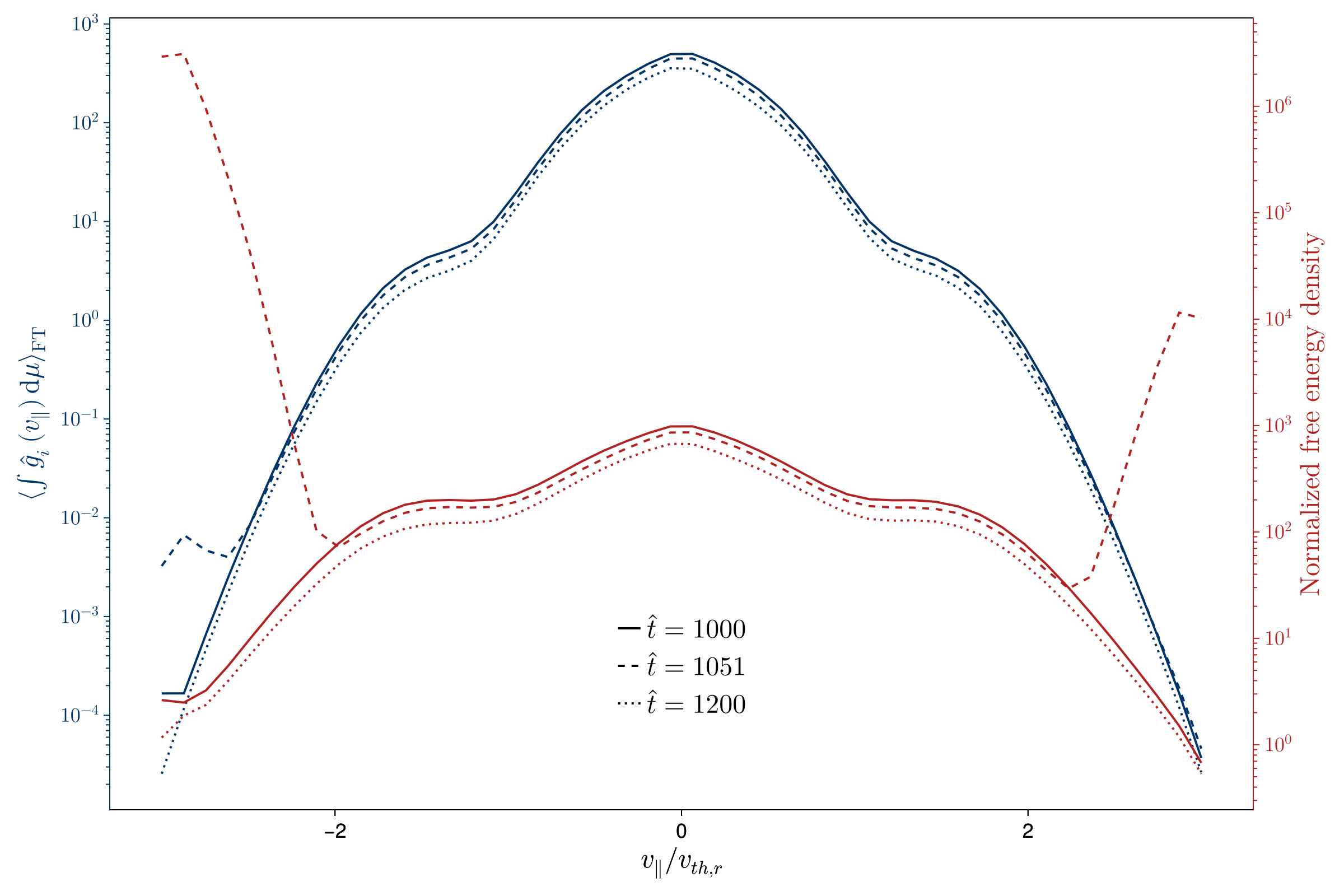}}
    \caption{Spectra of the distribution function and the normalised free energy as a function of \(v_\parallel\) at three different time steps, integrated over the flux tube and \(\mu\):  \(t = 1000\) (prior to the onset of the instability), \(t=1051\) (during the spike), and \(t=1200\) (following the instability).\label{fig:num_inst_vmu}}
\end{figure*}

To investigate the origin of these instabilities, we examine the distribution of the free energy in both spectral and velocity space. As shown in \Cref{fig:num_inst_kxkyz}, the instability manifests as a sudden accumulation of energy at the high-wavenumber edges of the spectrum , suggesting a spectral \enquote{pile-up} or aliasing effect. This effect is absent in the density fluctuations. Further analysis in velocity space, see \Cref{fig:num_inst_vmu}, reveals that these spikes are caused by an unphysical accumulation of free energy at the boundaries of the parallel velocity \(v_\parallel\). To some extent this can be seen in the distribution function itself but the effect is subdominant there. The accumulation is driven by the magnetic mirror term in the parallel dynamics.

This discrepancy arises from the different weighting of the velocity space integrals. The standard diagnostics are weighted by the background Maxwellian distribution; since the Maxwellian decays exponentially at large \(v_\parallel\), any numerical noise or instabilities at the boundaries of the velocity domain are heavily suppressed and smoothed. In the definition of the free energy, the non-adiabatic response is divided by the background Maxwellian. Consequently, any contribution of \(\delta f\) near the boundaries of the velocity space, where the Maxwellian is smallest, is subject to significant amplifications of the fluctuations. 

This demonstrates that the free energy acts as a high-sensitivity \enquote{early warning system} for numerical instability. It allows the user to detect unphysical energy growth at the boundaries of the phase space long before such instabilities manifest as macroscopic changes in the conventional diagnostics. Although the standard diagnostics were not affected in the current case, it is plausible that there will be other simulations where these effects are not negligible.

\subsection{Spatial distribution of free energy}
While the spectral analysis provides a global overview of the energy cascade, it inherently averages out the spatial intermittency of the turbulence. To gain a more granular understanding of the system, it is necessary to examine the distribution of these quantities in real space. As \texttt{stella} is based on a spectral method, the data are natively stored in Fourier space. Therefore, an inverse Fourier transformation was performed to reconstruct the physical spatial maps. This transition from spectral to real space allows for a more detailed investigation of the localized structures that drive the turbulent transport.

\begin{figure*}[htb]
    \centerline{\includegraphics[width=\linewidth]{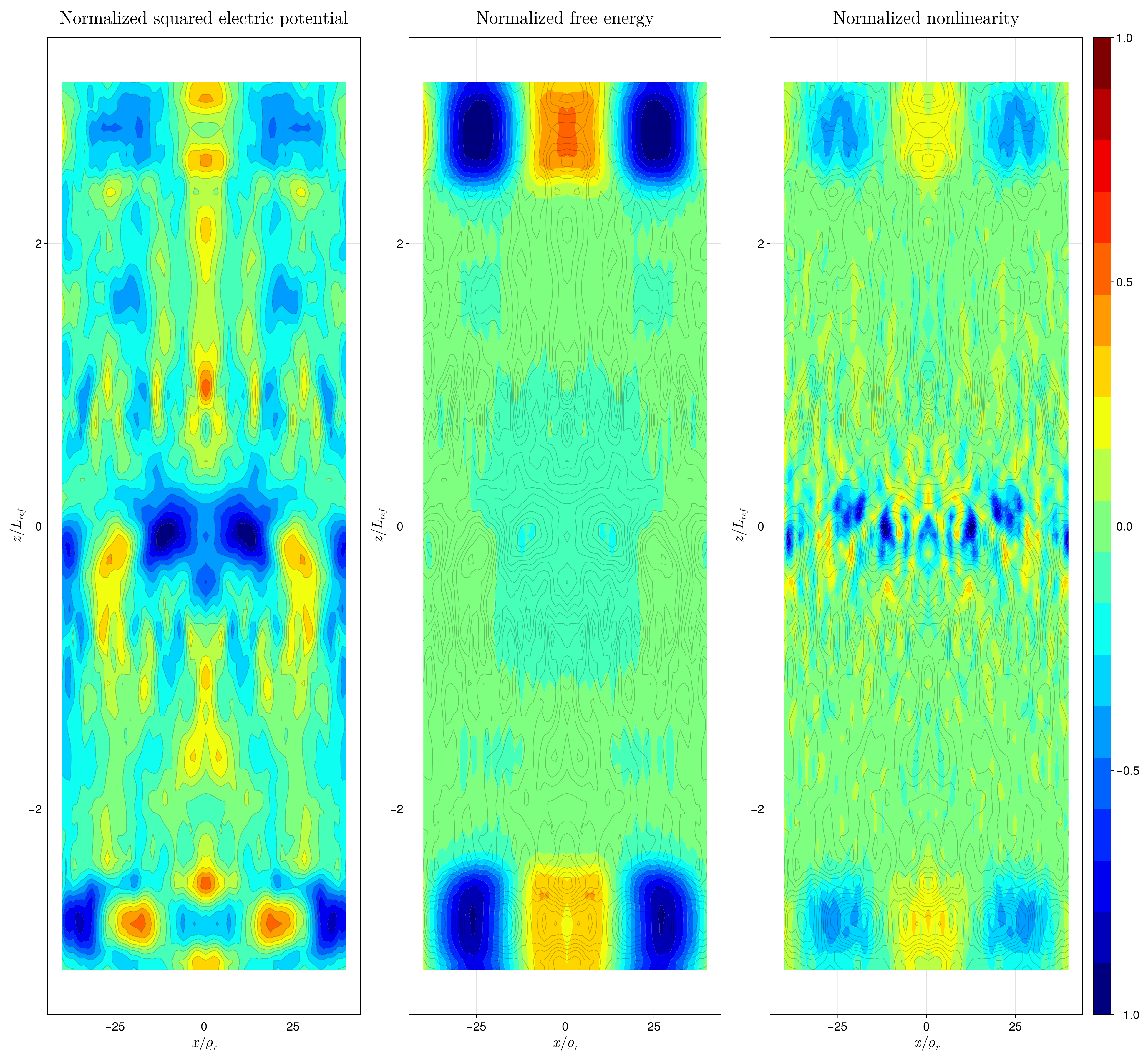}}
    \caption{Time-averaged spatial distributions of the squared electric field \(|\phi|^2\), the free energy, and the nonlinear term in the \((x,z)\) plane for the zonal mode \(k_y=0\). Black contour lines of \(|\phi|^2\) are overlaid on all panels to provide a common spatial reference. Results are shown for ITG turbulence at \(a/L_{T_i}=2.5\) and \(a/L_n=1.0\) for the high-mirror W7-X vacuum configuration, time-averaged over the saturated phase.\label{fig:spatial_dist}}
\end{figure*}

Drawing inspiration from the work of \cite{NavarroPoP2011}, which provided insights into the structure of the free energy in Fourier space, we extend this analysis to real space to identify features that may be less prominent in the electrostatic potential. In \Cref{fig:spatial_dist}, we compare the spatial maps of the squared electrostatic potential (\(|\phi|^2\)), the free energy, and the nonlinear free-energy transfer term. We extracted the zonal mode \(k_y=0\) and transformed it to real space in the \((x,z)\)-plane. To facilitate a direct comparison between the different quantities, the squared electrostatic potential is superimposed on all panels as black contour lines.

The comparison between these three panels reveals that the free-energy diagnostic provides additional insights. While the $|\phi|^2$ map shows the general layout of the electrostatic structures, the free-energy map exhibits a higher degree of localization, especially near the parallel boundaries. Because the free energy incorporates fluctuations in both the electric field and the particle distribution function\cite{SchekochihinTAJSS2009}, it highlights \enquote{hot spots} of energetic activity that are not always perfectly correlated with the peaks of the electric field.

This decoupling occurs because the kinetic contribution to the free energy dominates in regions where magnetic curvature is strongest, typically near $z \approx \pm \pi$, even when the electrostatic potential is relatively weak. Such behaviour reflects the inherently kinetic nature of ITG turbulence: the instability drive arises from the interplay between curvature-driven particle drifts and the resonant response of the distribution function, not from electrostatic effects alone. The free-energy diagnostic thus reveals regions of strong kinetic activity that would be overlooked by examining $\phi$ alone, highlighting the necessity of a full gyrokinetic invariant to pinpoint the actual sources of turbulent drive.

The comparison of the free-energy and nonlinear-transfer maps with the $|\phi|^2$ contours reveals a distinction between the large-scale organisation of the fluctuations and the regions of strong nonlinear
activity. Since the $k_y=0$ component is shown in the $(x,z)$ plane, the elongated structures in the parallel direction represent the coherent
parallel structure of the electrostatic fluctuations in this component. The $|\phi|^2$ contours show pronounced structures near the parallel boundaries and around the central region, providing a large-scale spatial reference for the turbulence.

The free-energy distribution follows some of this overall organisation, but is smoother and does not reproduce all of the fine structure visible in $|\phi|^2$. This indicates that the electrostatic amplitude alone does not determine the local distribution of free energy. The nonlinear term exhibits a more fine-scale and fragmented pattern, with enhanced activity around the central region and near the parallel boundaries. These structures identify regions where free energy is being actively redistributed rather than simply stored. The comparison therefore suggests that the coherent parallel structures provide the large-scale organisation of the fluctuations, while nonlinear transfer is concentrated in more localized regions within this structure. The free-energy diagnostic helps distinguish these two aspects by showing where energy is stored, whereas the nonlinear term indicates where it is redistributed.

The advantage of this spatial mapping is that it makes the separation between coherent large-scale organisation and localized nonlinear redistribution visible. While the electrostatic potential and free energy describe the spatial structure of the fluctuations and their stored energy, respectively, the nonlinear term highlights the more fragmented regions in which free energy is redistributed. This suggests that saturation is not spatially uniform, but involves localized nonlinear interactions that may be obscured by global spectral averages or by the electrostatic potential alone. Such information is useful for developing reduced transport models, for which global averages may not capture the locations of the strongest nonlinear interactions.
\section{Conclusion}

We have examined the numerical performance of several time integration schemes in the \texttt{stella} gyrokinetic code and demonstrated the utility of free energy as a diagnostic for both numerical stability and plasma turbulence dynamics. Linear stability analysis and nonlinear simulations revealed that the optimal scheme depends on the simulation regime: SSP22 achieves the lowest computational cost in linear ITG simulations, whereas SSP33 is most efficient in the nonlinear saturated state. The nonlinear simulations required approximately two to three orders of magnitude more computational time than the linear cases, underscoring the importance of efficient time integration for practical applications.

The free-energy diagnostic proved valuable beyond its formal role in gyrokinetic theory. It reveals characteristic power-law behaviour in the turbulent cascade, acts as a sensitive indicator of numerical instabilities that remain undetected in conventional diagnostics, and complements electrostatic-potential maps by distinguishing the spatial distribution of stored free energy from regions of localized nonlinear redistribution. These capabilities make free energy a useful framework for connecting numerical behaviour, turbulent energetics, and underlying plasma dynamics.

Our results provide practical guidance for selecting time-integration schemes in \texttt{stella} simulations and establish free-energy diagnostics as a valuable addition to the plasma-turbulence analysis toolkit. Future work will apply these diagnostics to quantify free-energy transfer between scales in W7-X turbulence and extend the scheme comparison to electromagnetic simulations and other magnetic configurations.


\subsection*{Acknowledgments}
The authors would like to thank the Max Planck Computing and Data Facility and the University of Greifswald for providing the computational resources used in this work.

\subsection*{Conflicts of Interest}
The authors declare no conflicts of interest.

\bibliographystyle{unsrt}
\bibliography{citations}
\end{document}